\documentclass[11pt,a4paper]{article}

\usepackage{lineno,hyperref}
\modulolinenumbers[5]
\usepackage{times}
\usepackage{amsmath}
\usepackage{balance}
\usepackage{amsfonts}
\usepackage{amssymb}
\usepackage{graphicx}
\usepackage{color}
\usepackage{enumerate}
\usepackage{color,soul,xspace}
\usepackage[tight]{subfigure}
\usepackage{float}
\usepackage{pdfpages}
\usepackage[table]{xcolor}

\usepackage[T1]{fontenc}
\usepackage[utf8]{inputenc}
\usepackage{authblk}

\newcommand*\samethanks[1][\value{footnote}]{\footnotemark[#1]}
\begin{document}


\title{Learning about Learning: Human Brain Sub-Network Biomarkers in fMRI Data}
\author[1]{Petko Bogdanov\thanks{Corresponding author.}\thanks{These authors contributed equally to the manuscript.}}
\author[2]{Nazli Dereli\samethanks}
\author[3,4]{Danielle S. Bassett}
\author[5]{Scott T. Grafton}
\author[2]{Ambuj K. Singh}
\affil[1]{Department of Computer Science, University at Albany, SUNY, 1400 Washington Ave, Albany, NY 12222, USA, pbogdanov@albany.edu. The research was performed while Petko Bogdanov was a postdoctoral researcher in Ambuj Singh's lab at UC Santa Barbara.}
\affil[2]{Department of Computer Science, University of California, Santa Barbara, CA 93106-5110, USA, \{ndereli,ambuj\}@cs.ucsb.edu}
\affil[3]{Complex Systems Group, Department of Bioengineering, University of Pennsylvania, Philadelphia, PA, 19104, USA, dsb@seas.upenn.edu}  
\affil[4]{Department of Electrical Engineering, University of Pennsylvania, Philadelphia, PA, 19104, USA, dsb@seas.upenn.edu}
\affil[5]{Department of Psychology and UCSB Brain Imaging Center, University of California, Santa Barbara, CA, USA, scott.grafton@psych.ucsb.edu}
\renewcommand\Authands{ and }

\date{}

\maketitle

\begin{abstract}
It has become increasingly popular to study the brain as a network due to the realization that functionality cannot be explained exclusively by independent activation of specialized regions. Instead, across a large spectrum of behaviors, function arises due to the dynamic interactions between brain regions. The existing literature on functional brain networks focuses mainly on a battery of network properties characterizing the ``resting state'' using for example the modularity, clustering, or path length among regions. In contrast, we seek to uncover subgraphs of functional connectivity that predict or drive individual differences in sensorimotor learning across subjects. We employ a principled approach for the discovery of significant subgraphs of functional connectivity, induced by brain activity (measured via fMRI imaging) while subjects perform a motor learning task. Our aim is to uncover patterns of functional connectivity that discriminate between high and low rates of learning among subjects. The discovery of such significant discriminative subgraphs promises a better data-driven understanding of the dynamic brain processes associated with brain plasticity.

\end{abstract}



\section{Introduction}
Network-based modeling and characterization of brain architectures has provided both a framework for integrating multi-modal imaging data as well as for understanding the function and dynamics of the brain and its subunits. Brain networks are traditionally constructed either from structural or functional imaging data. Structural brain networks represent properties of physical neuronal bundles and employ both invasive methods in animals~\cite{Markov2013} and non-invasive \emph{in vivo} methods such as diffusion MRI in humans~\cite{Park2004,Hagmann2007,Park2013}. Functional brain networks represent the functional associations between regions estimated by statistical similarities in regional time series, including correlation or coherence~\cite{Summerfield2006,Chun2011,Bassett2011}. Functional brain networks can be extracted from multiple types of neuroimaging data. In the case of fMRI data, regional gray matter activity is measured by the \textit{blood oxygenation level dependent (BOLD)} signal.  EEG or MEG data provide regional activity in the form of electrical activity and magnetic flux respectively.

Brain networks are commonly studied using techniques drawn from graph theory and machine learning~\cite{Turk-Browne2013}. These techniques provide fundamental and generalizable mathematical representations of complex neuroimaging data: nodes represent brain regions and edges represent structural or functional connectivity. This simplified graphical representation enables the principled examination of patterns of brain connectivity across cognitive and disease states~\cite{Stam2007}. While the majority of network-based studies have focused on the brain's default mode or ``resting'' state~\cite{Greicius2009}, more recent efforts have turned to understanding brain connectivity elicited by task demands, including visual processing~\cite{Summerfield2006,Chun2011} and learning~\cite{Bassett2011}. Global network analysis of both functional and structural connectivity has demonstrated that brain networks have characteristic topological properties, including dense modular structures and efficient long-distance paths~\cite{Bassett2006}.

However, traditional network analysis tools are not always sensitive to small perturbations in functional or structural connectivity. Recent efforts have focused on developing new tools to identify specific subgraphs that are particularly discriminative between brain states (cognitive or disease) and therefore critical for an understanding of local neurophysiological processes. Zalesky and colleagues describe a set of methods to identify groups of edges that are significantly different between two groups of networks~\cite{zalesky2010network,zalesky2012connectivity}. Motifs, defined as frequently occurring (across sessions and subjects) patterns of local connectivity, are groups of edges with particular topological properties that may play specific functional roles~\cite{echtermeyer2011automatic,kashtan2005spontaneous}. Finally, hyperedges are a type of edge set that may discriminate between cognitive states. Hyperedges can be defined as groups of edges that vary significantly in weight over time~\cite{bassett2014cross}, for example during adaptive functions like learning. In general, these tools seek to associate local network features (or subgraphs) with cognitive function, a critical step necessary for informing therapeutic interventions.

Here we develop and apply a novel method for identifying subgraphs that discriminate between individuals with differing behavioral variables. Drawing on new machine learning methods, we uncover the subgraphs that maximize the discriminative potential in explaining the differences in the rate of motor learning between individuals. The data for our analysis comes from a motor learning task experiment in which subjects' neural activity was measured using fMRI in multiple repeated sessions as they learned a set of new 12-note finger sequences analogous to piano arpeggios~\cite{Bassett2011}.

Based on the behavioral data, we assign individuals in the study to two categories -- ``high'' and ``low'' rate learners -- using the exponential rate of decrease in the time required to perform a motor sequence (also known as \emph{movement time}). The fMRI data was aligned to the Harvard-Oxford Brain Atlas (part of the FSL tool~\cite{jenkinson2012fsl}) involving $112$ cortical and subcortical regions. A functional edge strength linking two cortical areas was estimated as the wavelet-based coherence of the corresponding regional time series. 

Next, we employ and extend recent techniques from labeled network mining~\cite{MINDS} to uncover what we call \textit{biomarkers}---connected subgraphs of functional edges---whose strength of connectivity predict whether individuals are learning the task at ``high'' or ``low'' rate. More specifically, we binarize the weight of each edge at each time point into either high or low states, indicating strong or weak functional coherence between the associated brain areas. We then build in-network decision trees to guide the discovery of significant motor-learning-related subgraphs, whose edge state dynamics predict individual differences in learning rate. We employ randomized statistical tests to establish the significance of the discovered subgraphs and demonstrate their relationship to the subject's learning rates. Our work is the first to identify ``biomarkers'' of learning rate. Amongst tens of thousands of possible edges between $112$ cortical regions, we find functional subgraphs comprised of $6$-$10$ edges that predict whether a subject is a high or low rate learner. At the same time, we ensure that the obtained subgraphs are statistically significant, by demonstrating that they are unlikely to occur in a non-parametric null model.

\section{Subnetwork Biomarkers Discriminate Between High and Low Learning Rates}

Our goal is to detect a set of functional edges connecting cortical regions whose state (high/low coherence) can predict individual differences in learning rate. We expect that learning-related changes in functional connectivity will be located in coordinated neural circuits~\cite{Fries2005}, and we therefore restrict our attention to predictive edges which form a connected subgraph. As the rate of learning increases, some functional edges within a subgraph of interest will fall into a low coherence state (i.e. coherence between their adjacent regions' activation will approach 0), while others will move into a high coherence state. We seek to understand this dynamics and extract logical structures (in the form of decision trees) that predict the global behavioral state of the session: high/low rate of learning. We call the sugraphs and their corresponding decision trees \emph{learning rate biomarkers} (or just \emph{biomarkers}).

\noindent{\textit{Definition:}} A \emph{Biomarker} is a statistically significant connected subgraph of functional edges whose state (high \emph{vs.} low coherence) can collectively \emph{differentiate} between high and low learning rate of subjects.

The potential of a biomarker to differentiate between functional networks corresponding to low or high learning rate sessions is called \textit{discriminative power} (and the biomarker is called \textit{discriminative}). We employ a discriminative biomarker mining approach to analyze functional networks constructed from fMRI scans of 18 subjects performing a motor learning task over 3 learning sessions that occurred on 3 different days (see Methods and Data for details). The sessions were divided into low and high learning rate sessions based on the average reduction in movement time for completing the motor task. The goal of our analysis was to identify biomarkers (subgraphs) that were \textit{discriminative} of the session type (high \emph{vs.} low rate learners) and \textit{minimal}, while at the same time statistically significant. Minimal, in the context of biomarkers, refers to the requirement that no subgraph of a biomarker can enable a similar accuracy in discriminating low from high learning rate sessions.

\begin{figure}[H]
    \centering
    \subfigure[]{
       \includegraphics[width=0.5\textwidth]{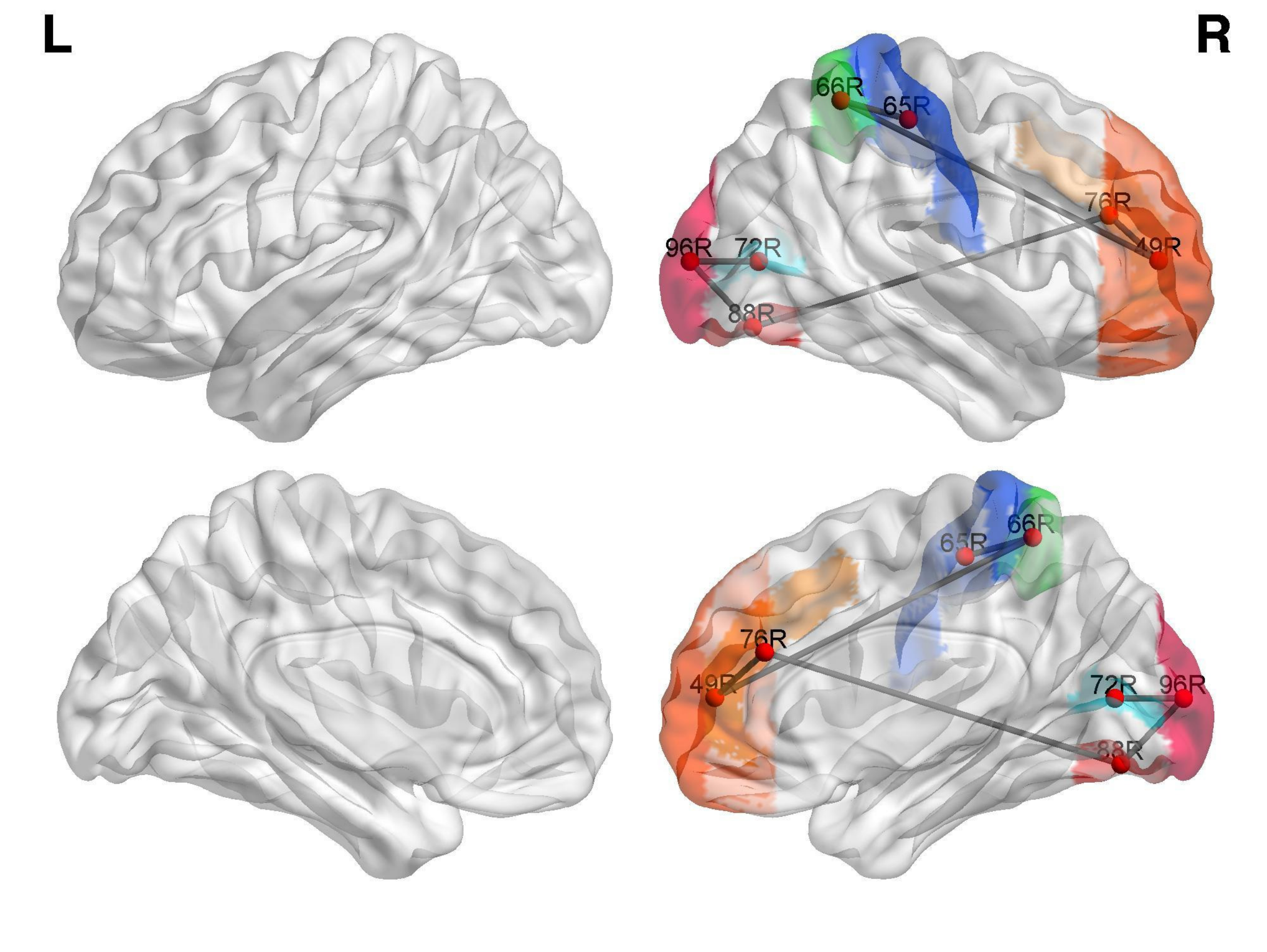}
       \label{fig:biomarker1}
    }\hspace{0.4in}
    \subfigure[] {
       \includegraphics[width=0.3\textwidth]{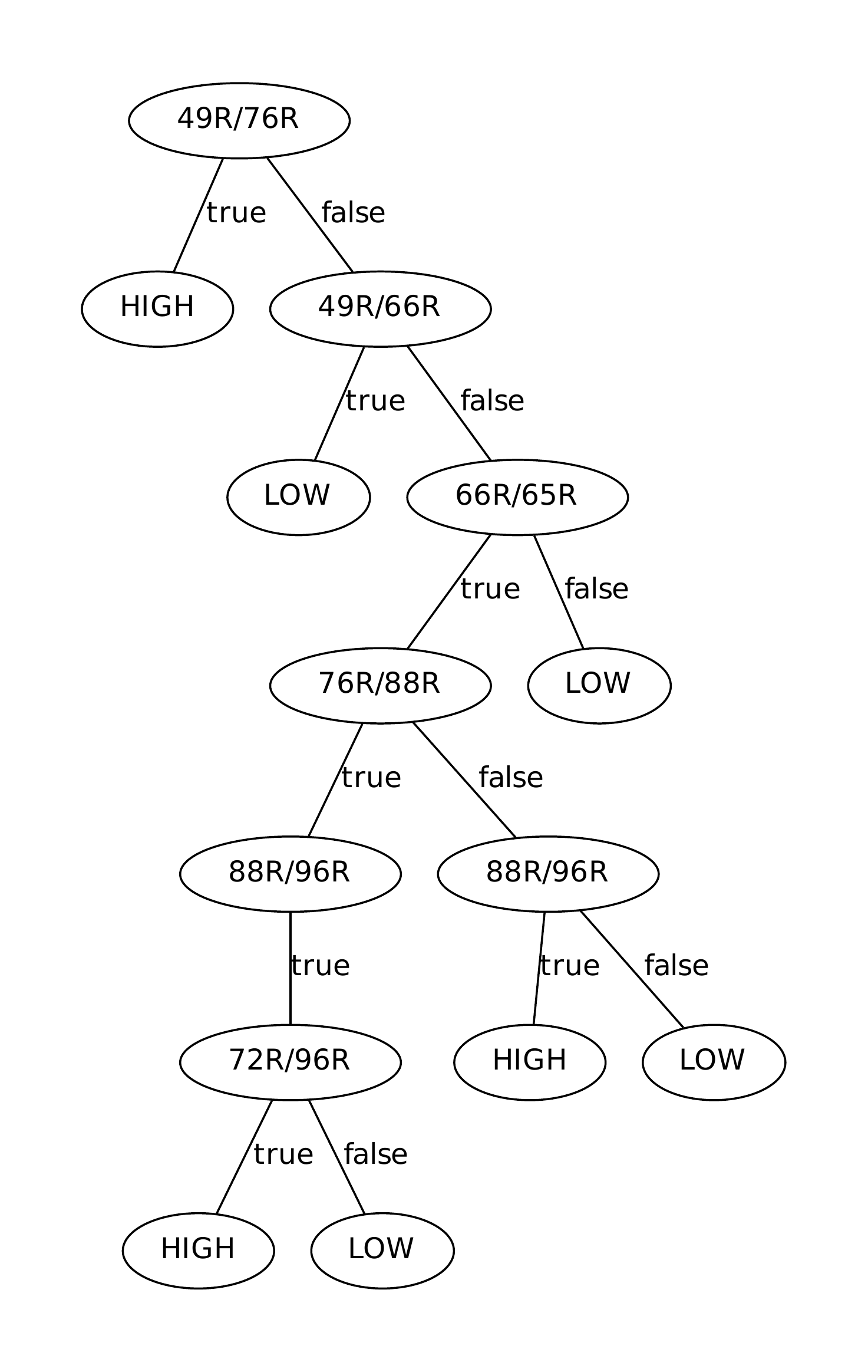}
       \label{fig:dectree1}
    }
    \subfigure[]{
       \includegraphics[width=0.5\textwidth]{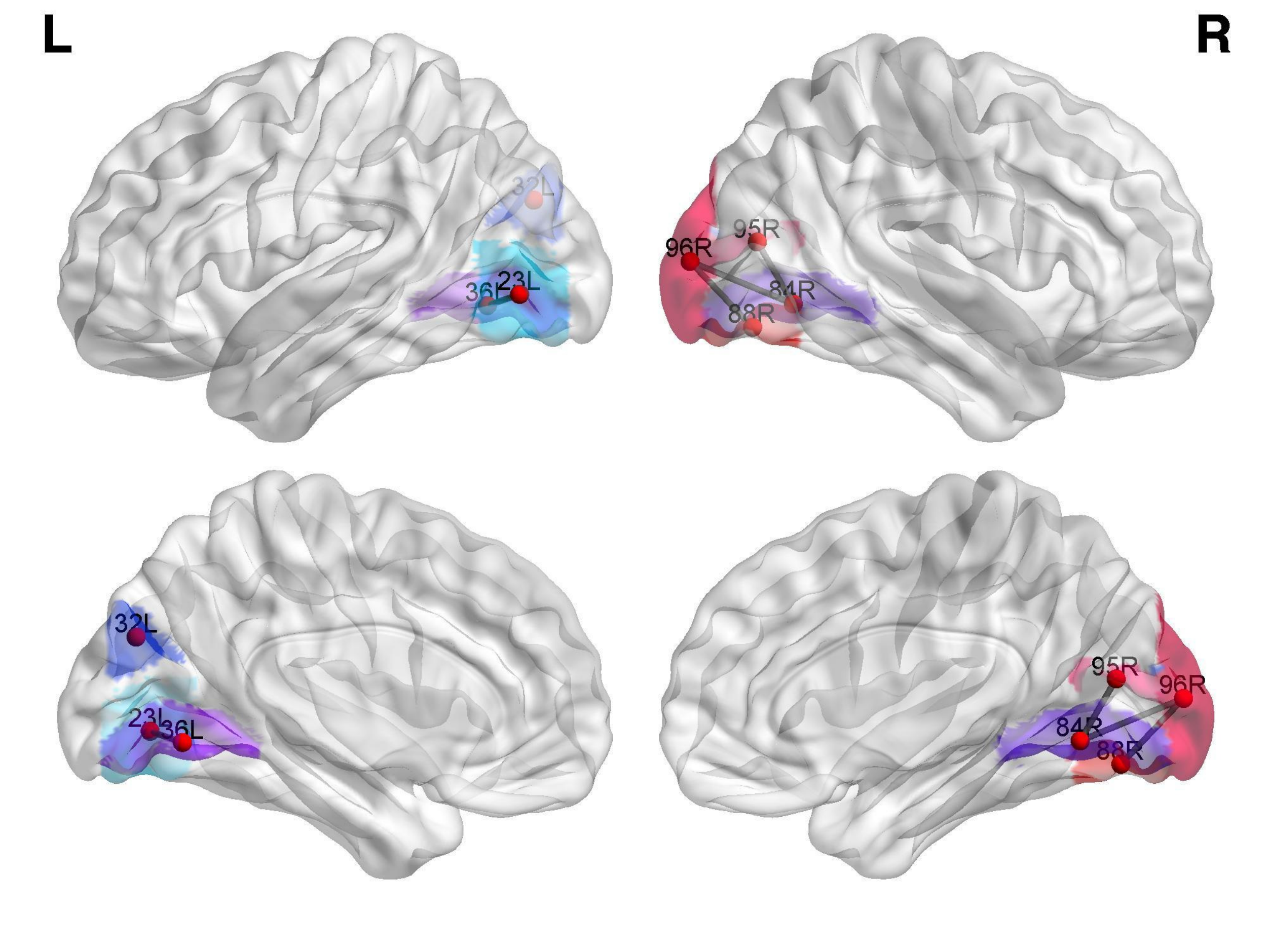}
       \label{fig:biomarker2}
    }\hspace{-0.3in}
    \subfigure[]{
       \includegraphics[width=0.5\textwidth]{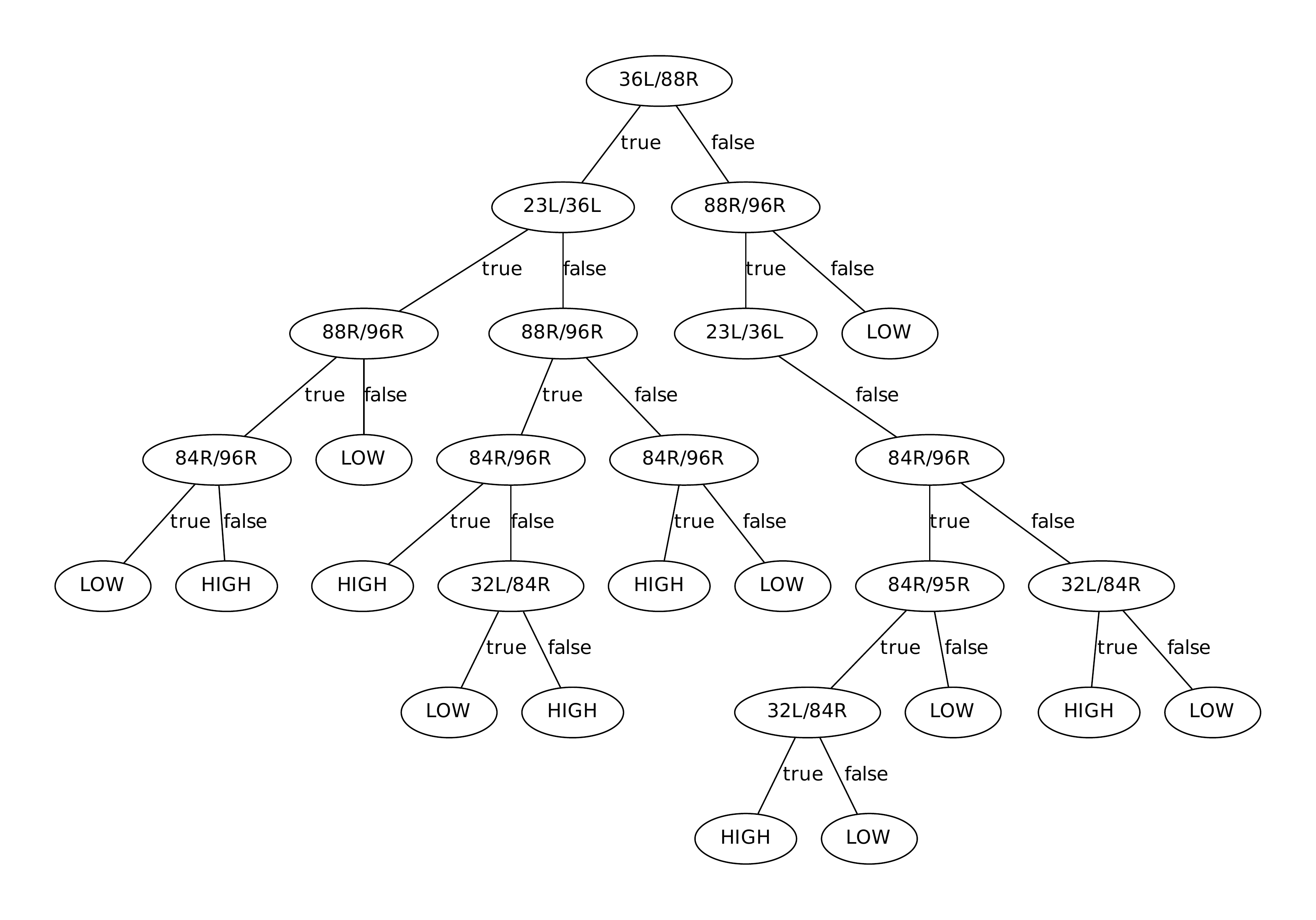}
       \label{fig:dectree2}
    }
    \caption{Two examples of discriminative biomarkers and their corresponding decision tree models. The decision tree \subref{fig:dectree1} corresponds to the biomarker in \subref{fig:biomarker1}.  According to one of its decision rules, a high learning rate session results from (1) a low-coherent edge between the right frontal pole (reg. 49) and the right paracingulate gyrus (region 76), (2) a low-coherent edge between the right frontal pole (reg. 49) and the right superior parietal lobule (reg. 66R), (3) a high-coherent edge between the right superior parietal lobule (reg. 66R) and the right postcentral gyrus (reg. 65), (4) a low-coherent edge between the right paracingulate gyrus (region 76) and the right occipital fusiform gyrus (region 88), and (5) a high-coherent edge between the right occipital fusiform gyrus (region 88) and the right occipital pole (region 96). Similar decision rules follow from any path from the root to a leaf of the decision tree and similarly for the biomarker and decision tree in Fig.\subref{fig:biomarker2}, \subref{fig:dectree2}. The mapping between the used identifiers and anatomical names of the brain regions is provided in the appendix.}
    \label{fig:example}
\end{figure}

In Figure~\ref{fig:example}, we show two example biomarkers and their corresponding \textit{decision trees}. Both biomarkers involve $7$ brain regions interconnected by $6$ functional edges (Fig.~\ref{fig:biomarker1} ,\ref{fig:biomarker2}). A decision tree in our context is a classification model that separates high from low learning rates within training sessions performed by 
subjects, based on the state of their observed functional edges (high or low coherence). The decision trees corresponding to our example biomarkers (Fig.~\ref{fig:dectree1} ,\ref{fig:dectree2}) both provide perfect ($100\%$) accuracy in predicting the learning rate. A path from the root to a leaf of the tree corresponds to a logical conjunctive decision rule involving the states (high/low coherence) of functional edges that implies a high/low learning rate in task performance.

The decision tree corresponding to the second biomarker example~Fig.~\ref{fig:biomarker2} is depicted in Fig.~\ref{fig:dectree2}. One possible decision rule from this tree predicts a high learning rate based on (1) a high-coherence state of the edge between the left lingual gyrus (region 36) and the right occipital fusiform gyrus (region 88), (2) a low-coherence state of the edge between the left lateral inferior occipital cortex (region 23) and the left lingual gyrus (region 36), (3) a low-coherence state of the edge between the right occipital fusiform gyrus (region 88) and the right occipital pole (region 96), and (4) a high-coherence state of the edge between right lingual gyrus (region 84) and the right occipital pole (region 96).

Our analysis based on discriminative biomarkers involves the following components:
\begin{enumerate}
\item We establish the \emph{significance of mined biomarkers}. To this end, we perform non-parametric permutation tests that reassign learning rates and edge states uniformly at random within session-specific networks. We compute the associated $p$-value of the number of true discovered biomarkers based on the distribution of number of discriminative small subgraphs in randomly permuted data.
\item We report significant brain areas and functional edges between them that occur frequently in mined biomarkers.
\item We visualize and discuss discovered biomarkers in the context of motor learning theory.
\end{enumerate}

\subsection{Discovering Significant Discriminative Subgraphs}

Biomarkers are significant subgraphs that can accurately discriminate between high and low learning rates of subjects. The space of potential biomarkers encompasses the set of all possible connected subgraphs of the functional brain network, which is exponential in the number of brain areas. Hence, the fair consideration of all subgraphs is computationally intractable even at the spatial resolution of 112 cortical and subcortical regions. Instead, we develop and employ a sampling approach called \textit{MINDS-prune} that extends a recent network-constrained decision tree inference approach called MINDS~\cite{MINDS}. The general idea behind MINDS-prune is to use a Markov Chain Monte Carlo (MCMC) sampling to efficiently extract predictive subgraphs. 

Employing a sampling technique has several implications: (i) it allows us to explore multiple biomarkers of high predictive power (ii) in a non-deterministic fashion and (iii) in a computationally efficient manner. Due to the non-deterministic nature of the approach, its output varies across executions and hence we perform multiple runs initiated by different starting subgraphs. In addition, we need to establish the statistical significance of subgraphs in comparison to an appropriate non-parametric null model. To this end, we perform two permutation-based tests in which either the state of a functional edge (high or low coherence) or the learning state (high or low rate) is permuted uniformly at random. We refer to these two tests as edge permutation and learning rate permutation, respectively. We compute the number of small (varying the maximum size between 4 and 14 edges) and discriminative (predictive accuracy exceeding 90\%) subgraphs and test the probability of observing the same or higher number of such subgraphs under our edge permutation (EP) and learning rate permutation (LP) null models (right $p$-values).

The input to our analysis is a set of $27$ coherence functional networks acquired from $18$ subjects over $3$ experimental sessions. In these networks, nodes correspond to cortical and subcortical regions and edges quantify the level of coherence for a pair of regions. The global state (label) of each network corresponds to either high or low learning rate. Note, that we disregard sessions preceded by high learning rate sessions for the same subject, since we expect that once the motor task sequence is learned (a high-learning rate session has been observed) subsequent sessions are not informative (see Methods and Data section for details).

We sample subgraphs common to all networks using MINDS-prune. MINDS-prune is an instance of Metropolis Hastings MCMC sampling. It proceeds by drawing a sequence of samples by proposing local moves from a given point in the state space. In the case of subgraphs, the current state is an instance of a subgraph and a local move involves addition or removal of a single node without violating the subgraph connectivity. A move may be rejected or accepted (depending on the change in classification accuracy), and in the case of the latter, the new state (subgraph) is added to the set of samples. To initiate the sampling, we pick uniformly at random a node and 10 neighbors of this node also chosen uniformly at random; possible neighbors are identified using a breadth-first-search algorithm. We use an initial ``burn-in'' period of $1000$ moves during which samples are discarded (not added to the sample set), a common practice in MCMC-style algorithms to reduce the dependence on the initial state. We than draw $2000$ sample subgraphs and apply a post-processing phase that retains only high-accuracy and minimal subgraphs. We analyze and discuss the selection of the above parameters at the end of our Methods section.

To adequately ``comb'' the search space, we perform $20$ independent runs of MINDS-prune, using the same parameter settings but different starting state subgraphs to initiate the process. Using the state of subgraph edges (high or low coherence) as features, we measure the predictive power of each subgraph based on its accuracy in classifying a session as high versus low learning rate. In this search, we retain subgraphs whose accuracy exceeds $90\%$, i.e. a subgraph is retained only if its corresponding decision tree classifies accurately at least $90\%$ of the sessions. Since we are working with a total of $27$ sessions, we effectively retain all subgraphs whose decision trees do not mis-classify more than two sessions. We allow for a few mis-classified examples so as to reduce the possibility of overfitting to the training set. We also test the generalization (lack of overfitting) of our approach using cross validation in Sec.~\ref{sec:mining}.

\begin{table}[h]
\centering
{\scriptsize
\begin{tabular}{|l|c|c|c|} \hline
 Max Size    & \# Biomarkers & LP p-value & EP p-value \\ \hline	
$4$           &$1$             &$0.38$    &$<0.01$\\ \hline
$5$           &$3$             &$0.33$    &$<0.01$\\ \hline
$6$           &$7$             &$0.18$    &$<0.01$\\ \hline
$7$           &$16$            &$0.05$    &$<0.01$\\ \hline
$8$           &$24$            &$\mathbf{0.03}$    &$\mathbf{<0.01}$\\ \hline
$9$           &$30$            &$\mathbf{0.04}$    &$\mathbf{<0.01}$\\ \hline
$10$           &$34$            &$\mathbf{0.03}$    &$\mathbf{<0.01}$\\ \hline
$11$           &$37$            &$\mathbf{0.02}$    &$\mathbf{<0.01}$\\ \hline
$12$           &$38$            &$\mathbf{0.03}$    &$\mathbf{<0.01}$\\ \hline
$13$           &$39$            &$\mathbf{0.02}$    &$\mathbf{<0.01}$\\ \hline
$14$           &$40$            &$\mathbf{0.03}$    &$\mathbf{<0.01}$\\ \hline
$15$           &$40$            &$\mathbf{0.04}$    &$\mathbf{<0.01}$\\ \hline
$16$           &$40$            &$0.08$    &$0.04$\\ \hline
$17$           &$40$            &$0.11$    &$0.08$\\ \hline

\end{tabular}	
}
\caption{Average number of accurate (accuracy of predicting low from high learning rate sessions exceeding $90\%$) biomarkers of increasing maximal size. The significance of observing the obtained observed number of accurate biomarkers is tested based on two kinds of permutation tests: learning rate permutation (LP) in which the session learning rates are uniformly permuted; and edge permutation (EP) in which the high/low coherence states of functional links within networks are uniformly permuted. The $p$-values of the observations are presented in the third and fourth columns.}
\label{tab:pvalue}
\end{table}

To evaluate the significance of the obtained decision trees, we perform two types of non-parametric permutation tests: (i) a permutation of learning rate labels (high versus low learning rate) and (ii) a permutation of edge states (high versus low coherence). For both tests, we measure the probability of obtaining similar-size trees of high accuracy. For the learning rate permutation test (LP), we shuffle the high versus low learning rate labels, while keeping their fractions in the training set of sessions unchanged. For the edge state permutation test (EP), we shuffle the high versus low edge coherence labels, while keeping their fraction constant in each session-specific functional network. We perform $100$ permutations of each test. Then, we perform (similar to the protocol for the observed data) $20$ runs of MINDS-prune for each permuted instance and retain subgraphs whose accuracy exceeds $90\%$. We expect that neurophysiologically relevant subgraphs will provide high accuracy with only a few edges, whereas subgraphs obtained from permuted data will provide comparable or lower accuracy with many edges. Hence, we compute the right-tailed $p$-value of the average number of unique subgraphs of size at most $k$ edges, using the obtained subgraph sizes in permuted networks to derive a background distribution.

In Table~\ref{tab:pvalue}, we show the number of unique high-accuracy subgraphs and the corresponding $p$-values of observing as many high-accuracy subgraphs in the two different permutation background distributions. While the MCMC process may traverse the same subgraph more than once, we disregard such multiplicity information and measure the significance of the number of unique accurate trees. Multiple samples of the same subgraph do not necessarily give evidence for its quality as the MCMC process may spend more time near a local optimum. In addition, the significance of the number of unique accurate trees has a more intuitive interpretation as opposed to counting repeated occurrences.

As evident in Table~\ref{tab:pvalue}, on average we observe few accurate subgraphs of size $7$ or smaller, and their observation is not significant in the context of random learning rate permutation LP ($p\geq0.05$). Larger accurate subgraphs (sizes 8-14 edges) are more numerous, and their occurrence is significant based on both the LP and EP background distribution tests (unadjusted $p\leq0.04$ for both random tests and $p\leq0.02$ for size 11). Larger accurate subgraphs (size 16 and higher) become insignificant according to both randomization tests. The above analysis, suggests that a coherence level of up to $10$ functional edges among cortical regions (out of approximately $6000$ at the spatial resolution of 112 regions) are sufficient to differentiate between high and low learning rate sessions. Next, we discuss the edges and areas that consistently occur in the discriminative biomarkers.

\subsection{Brain Areas and Functional Edges Consistently Identified in Biomarkers}

After mining discriminative biomarkers, we ask whether specific brain areas or functional edges are consistently identified as discriminative. For this analysis, we focus on accurate biomarkers of at most $10$ edges. This cut-off size results in a significant number of accurate subgraphs discovered (p-value $\leq 0.03$) as compared to other cut-off sizes (Table~\ref{tab:pvalue}). Simultaneously, the subgraphs are relatively small, involving less than $10\%$ of all 112 brain areas). Changing the maximal size of the subgraph while maintaining statistical significance (e.g. sizes 8-14 edges) does not have a qualitative effect on the set of regions or the set of edges that are frequently identified. We demonstrate this stability at the end of this section.

We compare the observed frequencies of occurrence of regions and functional edges in biomarkers to their corresponding frequency distributions under the random edge permutation model (EP discussed in the previous section). We quantify the statistical significance of the observed frequencies by calculating their associated $p$-values. Since, we compare the real and permuted data in terms of multiple region frequencies, we need to correct for multiple comparisons. We control for the false discovery rate (FDR) at a level of $0.05$ according to the Benjamini-Hochberg step-up procedure~\cite{benjamini1995controlling}. According to the latter, the hypotheses (biomarker regions in our case) are sorted by their $p$-values in ascending order. Regions corresponding to indices up to $k$ are declared significant, where $k$ is the maximal index such that $p(k)\leq (k/n)\alpha$, $p(k)$ is the corresponding $p$-value, $n$ is the total number of regions (112), and $\alpha $ is the FDR level set to $0.05$.

\begin{figure}[t]
\centering
\label{fig:significant_regions}
\includegraphics[width=0.9\textwidth]{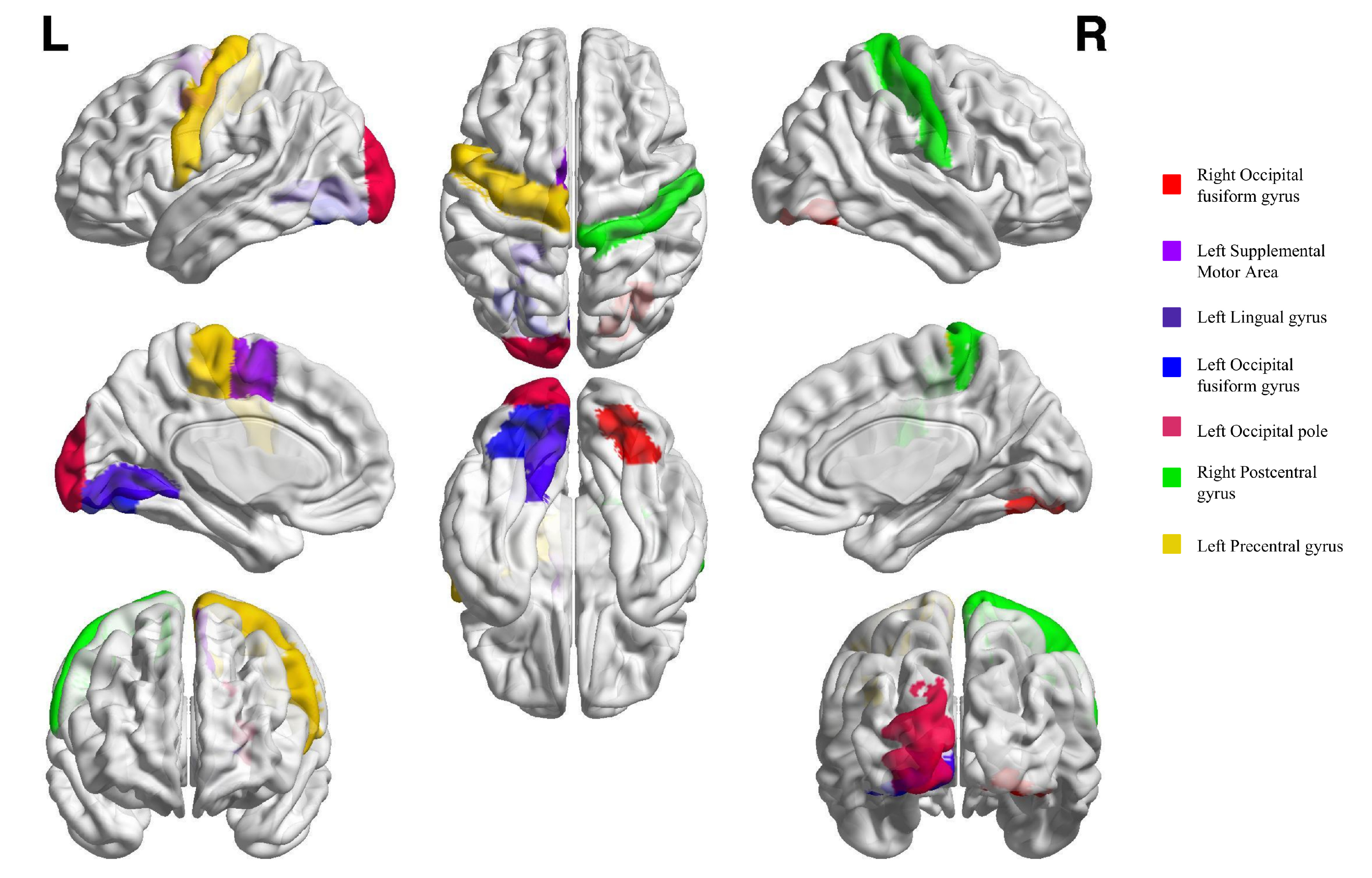}
\caption{Brain regions significantly more represented in the mined biomarkers than expected in permutation distributions. These areas are predominantly located in the motor and visual cortices.}
\end{figure}

The most frequent significant regions (depicted in Figure~2
) in biomarkers are located in the motor and visual cortices, and include the \textit{occipital fusiform gyrus, left supplemental motor area, left lingual gyrus, left occipital pole, right postcentral gyrus} and \textit{left precentral gyrus}. Most of these regions play an essential role in visually-guided movement. For example, the fusiform gyrus contains higher-order visual association areas that control spatial vision and attention processes~\cite{orrison2008atlas}. The involvement of these regions is parsimonious with the functional requirements of the particular learning task used in this experiment. Subjects were required to play out twelve key presses as indicated on a 4-staff notation system. This requires subjects to learn to both read a novel visual notation and to remember particular visual patterns that indicate the sequential key presses.

Apart from vision, the regional results are noteworthy for the involvement of precentral gyrus (motor cortex) and supplementary motor area along with the associated somatosensory cortex. The primary and supplementary motor cortex are known to be involved in the sequential control of movement and often show changes over time with sequence learning~\cite{grafton1995functional,hazeltine1997attention,bischoff2004neural,wymbs2013contributions}.  Together, these results make a clear case that fast learners are better able to leverage visiomotor areas to generate perceptual as well as movement related representations of the sequences. More generally, it suggests that, not surprisingly, successful learning involves the recruitment of appropriate brain systems to represent information that is relevant for sculpting behavior.

Note that our method discovers these areas in an unsupervised manner (in terms of known area functions), using only information from fMRI and measurements of learning rates in the sensorimotor task. The alignment of discriminative regions to areas known to be involved in motor learning supports the utility of our approach and its potential applicability in other cognitive domains.

\begin{figure}[t]
\centering
\includegraphics[width=0.9\textwidth]{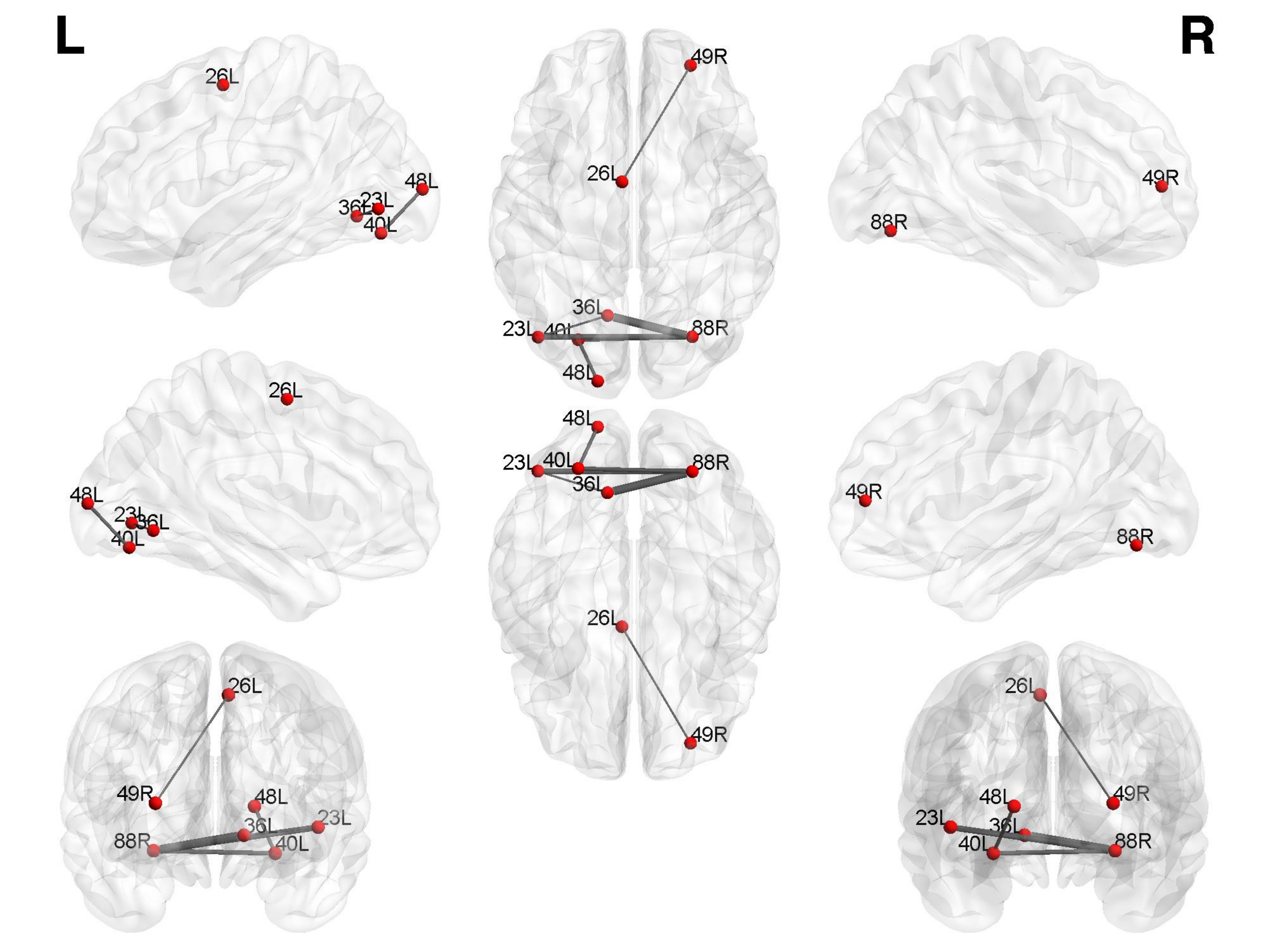}
\caption{Significantly frequent functional edges found in the mined biomarkers. The thickness of the edge is proportional to its frequency. Solidifying the findings from the analysis at the level of regions (Fig.~\ref{fig:significant_regions}), significant functional edges interconnect motor and visual cortex regions.}
\label{fig:edges}
\end{figure}

Beyond individual cortical or subcortical regions, we also consider \emph{edges} that occur more frequently in biomarkers than expected in the random edge permutation null model. We similarly correct for FDR at a level of $0.05$. In this case, the number of hypotheses involved in the Benjamini-Hochberg correction procedure is the number of all possible functional edges: $6216$.  In Figure~\ref{fig:edges}, we show the significant most frequent edges. Consistent with the regional findings, these edges tend to connect areas of the motor and visual systems.

The frequently occurring connections between the frontal pole and supplemental motor area are intriguing. Given the role of the frontal cortex in executive control, sustained attention to a task and the organization of sequential behavior, we can speculate that these different functional roles are interacting with the actual control of movement in this task. It is possible that an ability to enhance executive control during the task might facilitate the sustained recruitment of motor areas, keeping the subjects ``on-task''. This in turn would enhance the rate of learning. Similarly the numerous connections involving the occipital fusiform gyrus and the visual cortex is another interesting pattern. The most frequent connection involved in biomarkers relates the lingual gyrus and occipital fusiform gyrus which are both involved in visual processing and their functional interaction is probably needed to encode the visual cues that define a sequence. Recognizing a particular visual sequence can greatly enhance the ability to plan and execute a corresponding motor sequence. This is an obvious reason why the recruitment of these functional edges could accelerate learning.

\begin{figure}[H]
    \centering
    \label{fig:stability}
    \subfigure[Stability of top regions]{
       \includegraphics[width=0.4\textwidth]{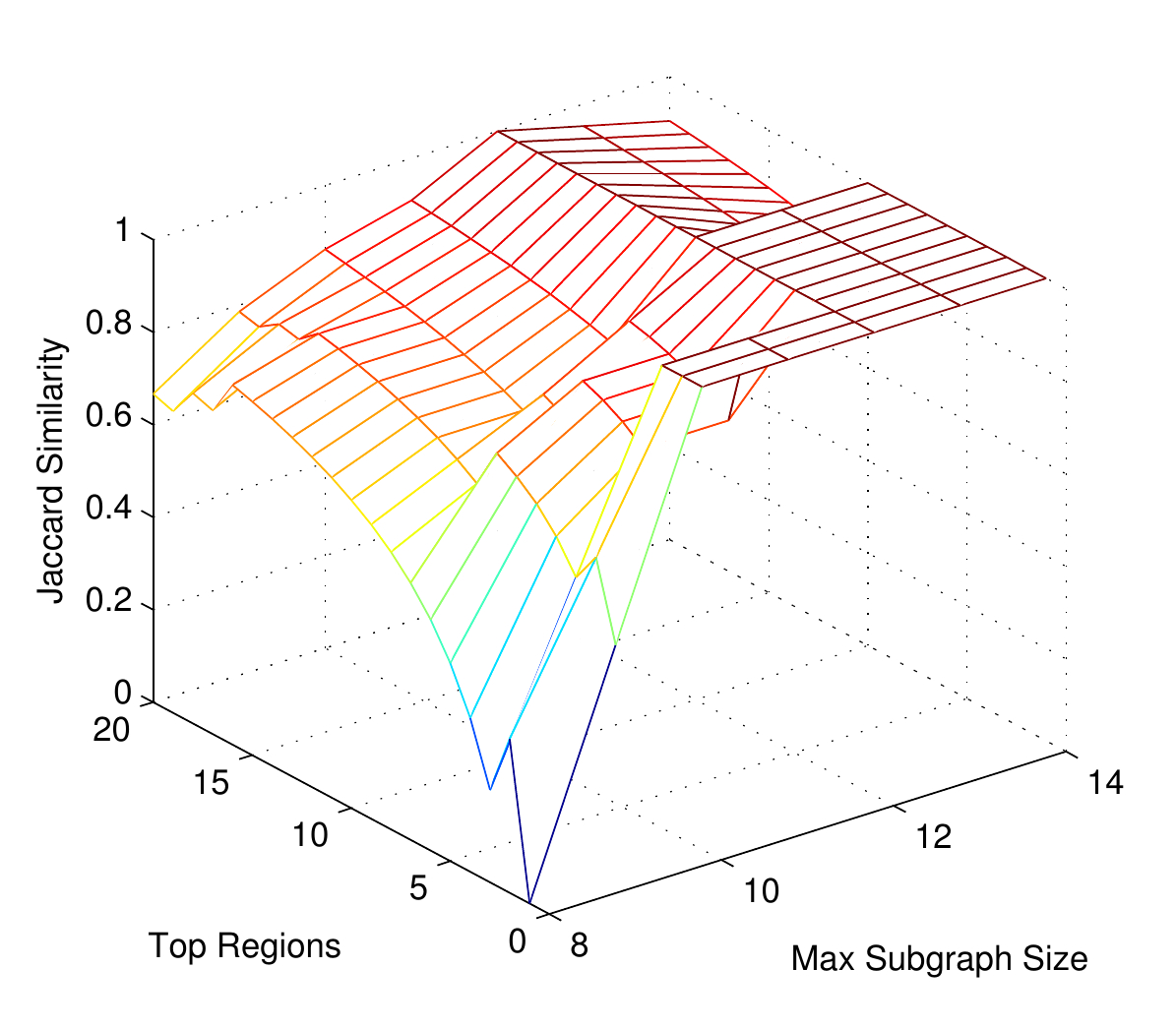}
       \label{fig:nodes_stability}
    }
    \subfigure[Stability of top edges] {
       \includegraphics[width=0.4\textwidth]{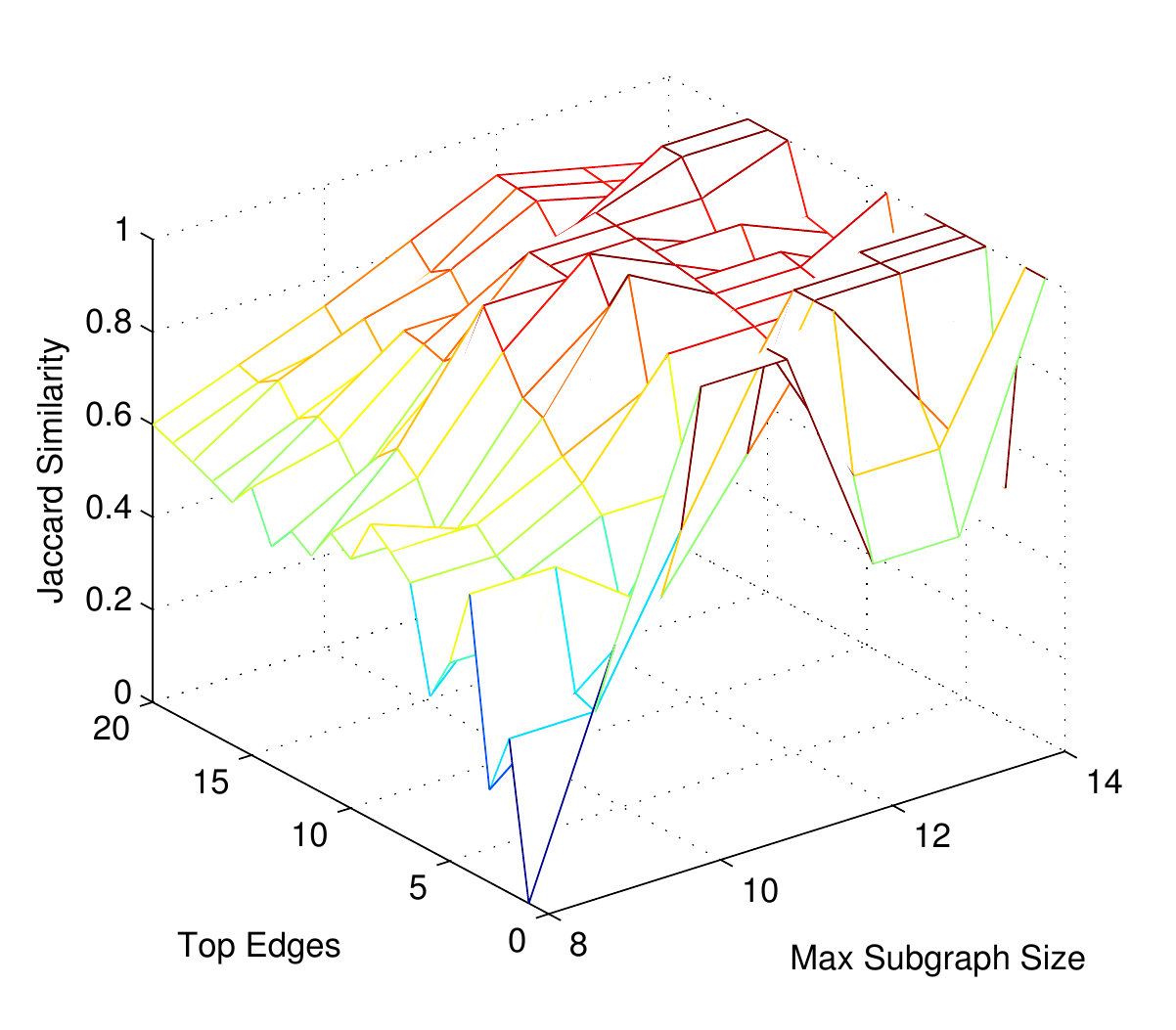}
       \label{fig:edges_stability}
    }
    \caption{Local similarity of top region and edge sets when varying the maximal size of discriminative subgraphs. The average Jaccard similarity of the top nodes/edges at a given maximal size and its neighboring sizes is reported on the vertical axis. The set of top regions and edges is stable (local similarity close to 1) for max subgraph sizes of 10-14. We focus on maximal sizes 8-14 because the number of mined subgraphs is significant for those values.}
\end{figure}

To test the sensitivity of the frequent regions and edges to the setting of the maximal subgraph size, we vary the maximal subgraph size and compute the similarity of the top region/edge sets to the same when considering one-edge smaller or larger maximal sizes (Figs.~\ref{fig:nodes_stability},\ref{fig:edges_stability}). As long as subgraphs of 10 edges are considered, the sets of top regions and edges stabilize (Jaccard similarity close to 1). Sizes higher than 14 become less significant (see Tab.~\ref{tab:pvalue}.)

\section{Methods and Data}

\subsection{Experimental Data from Motor Learning}
The data for our analysis was collected during a motor learning task experiment in which subjects' neural activity was measured using fMRI~\cite{Bassett2011}. The data was originally used to analyze the brain's functional flexibility during learning. Here, we follow the same protocol for data preparation, but focus on subgraph biomarkers associated with learning. Next, we shortly summarize the performed learning task experiment and data preparation. For more details, we refer the reader to the original paper introducing the experiment and analyses by Bassett and colleagues~\cite{Bassett2011,wymbs2012}.

The study involved 18 paid participants without formal training in playing a musical instrument, with normal vision, and without neurological or psychiatric disorders. In the motor learning task, subjects responded to a visually cued sequence by generating responses using the four fingers of their non-dominant hand on a custom-built response box. Visual cues were presented as a series of dots on lines, like musical notes, on a pseudo-musical staff with each line corresponding to one of the buttons on the response box to be pressed. The number and order of sequence trials was identical for all participants. All participants completed three training sessions in a five-day period, and each session was performed inside the MRI scanner (3.0 T Siemens Trio with a 12-channel phased-array head coil).

We used the Harvard-Oxford Atlas (included the FSL tool~\cite{jenkinson2012fsl}) that partitions the brain into 112 cortical and subcortical structures for aligning the fMRI images. A regional activation level time series was estimated based on averaging voxel intensities (details of imaging acquisition and normalization are available in~\cite{Bassett2011}). Coherence for all pairs of regional time series was then computed and subjected to statistical testing, preserving only values that passed a false discovery rate correction for multiple comparisons.

\subsection{From fMRI Data to Coherence Graphs and Biomarkers}

\begin{figure}
\centering
\includegraphics[width=0.9\textwidth]{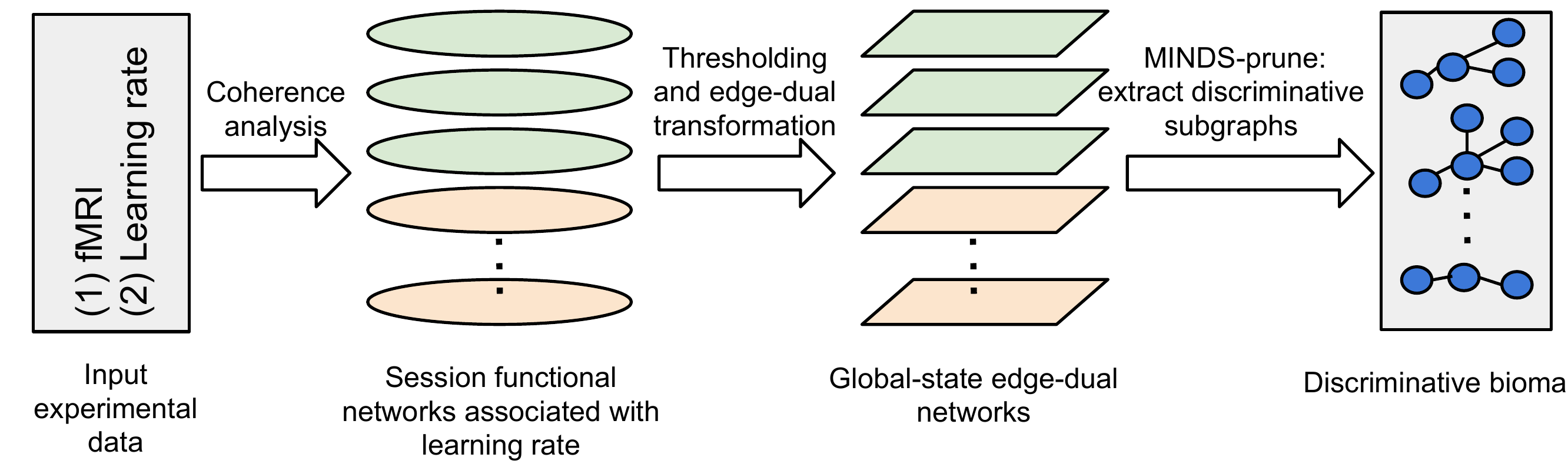}
\caption{An overview of our method. We start with experimental fMRI and learning rate measurements from individual sessions. We create coherence networks based on the fMRI for all sessions. The coherence networks are transformed into edge-dual graphs by binarizing the coherence states and edge learning rates. As a result, we obtain global state networks for sessions that are not preceded by high learning rate sessions, where the global leaning rate state is viewed as a class label of a network instance. We next employ MINDS-prune to extract discriminative and minimal subgraphs that can accurately predict the global state of all network instances.}
\label{fig:overview}
\end{figure}

Figure~\ref{fig:overview} provides an overview of our approach that transforms the input fMRI and learning rate data associated with sessions into a global-state network classification instance. We then extend the technique by Ranu and colleagues~\cite{MINDS} in order to sample discriminative and minimally connected subgraphs that predict the learning state in a session. In the remainder of this section we discuss the specifics of each of these steps.

A \textit{global-state network} is a graph with local labels on nodes and a global network state indicating the occurrence of an event (or the network type)~\cite{MINDS}. The goal in global state network classification and feature selection is to find small connected subgraphs involving the most discriminative nodes, whose labels predict the global state of network instances in a dataset. In our case, each training session completed by a subject corresponds to a \textit{functional network} that links brain regions based on coherence between regional time series. Session-specific network states correspond to high or low rate learning. Edge state labels indicate high or low coherence of the region time series.

In order to transform the data from the motor learning experiment discussed in the previous section to an instance of global network state classification, we (i) convert the  functional networks derived from training sessions into their edge-dual equivalents and then (ii) quantize both the local edge coherence values and global learning rate measurements into binary labels.

\begin{figure}[H]
\centering
\includegraphics[width=0.7\textwidth]{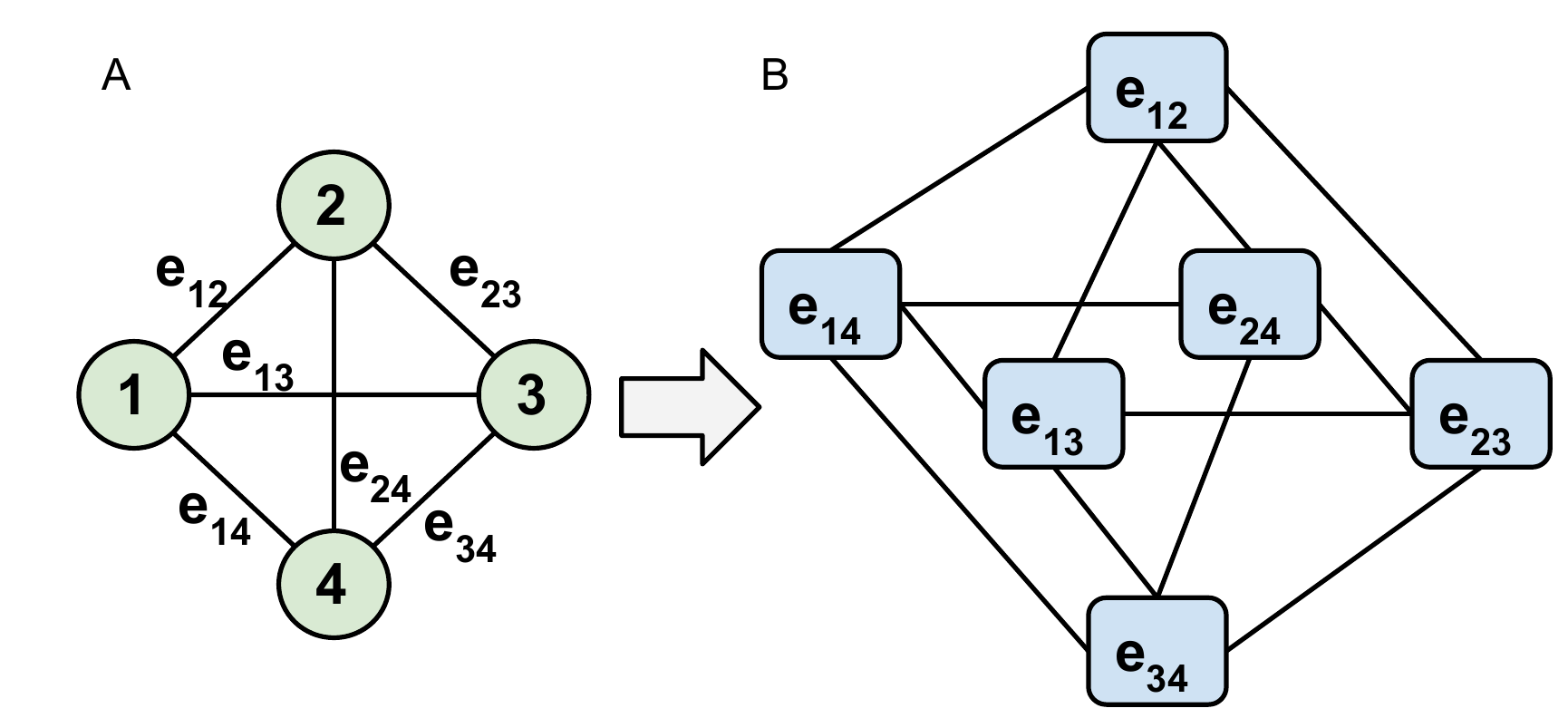}
\caption{Transformation from edge-weighted graph (A) to a corresponding edge-dual graph (B). Edges become nodes and are connected if they share a node in the original graph.}
\label{fig:transformation}
\end{figure}

\noindent{\bf Edge-Dual Functional Network.} We are interested in using the level of coherence of functional edges as features to inform the identification of discriminative biomarkers. However, the existing methods in global network state classification including MINDS~\cite{MINDS} work with node labels as opposed to edge labels. To convert our problem into this common framework, we transform the original functional network into its edge dual graph. Original functional edges become vertices in the edge-dual graph. Two vertices have a link between them if their corresponding edges in the original network share a common end node (region in the brain). To avoid confusion, we will use edges and nodes when discussing the original functional graph among brain regions (112 nodes and 6216 edges); vertices and links to refer to the edge-dual graph's elements (6216 vertices and 344988 links).

The transformation is demonstrated in Figure~\ref{fig:transformation} for a small example network of 4 nodes and 6 edges that gets transformed into its edge-dual. If we start with a complete graph $G(N,E)$ of $|N|$ nodes and $|E|=|N|(|N|-1)/2$ edges, the corresponding edge-dual graph $G_{ed}(E,L)$ will have $|E|$ vertices and $|L|=|E|(|N|-2)$ links. In our example graph, (Fig.~\ref{fig:transformation}) we have 4 nodes and 6 edges and in the corresponding edge-dual graph we obtain 6 vertices and 12 links. In this transformation, originally adjacent edges become nodes in the dual graph that are also connected by a link. Note, that this transformation ensures that a subgraph of connected vertices in the dual graph corresponds to a connected subgraph of edges in the original functional network. Next, we discuss how we binarize global networks states (learning rates) and local vertex states (coherence level).

\noindent{\bf Thresholding Coherence Values.} We threshold the vertex values in the edge-dual graph to obtain a binary feature corresponding to a vertex being in a high or low coherence state (note that the vertices correspond to the original functional edges). We choose a coherence value of $0.4$ as the threshold between the low and high coherence states. Effectively, $2.68\%$ of all nodes in functional networks are labeled as being in a high coherence state. To establish this threshold value, we experimented with lower and higher alternatives and examined the corresponding mined subgraphs. The latter were either insignificant under the learning state permutation (LP) test (for lower thresholds), or the classification accuracy of the obtained subgraphs decreased (for higher thresholds) since almost all nodes are reduced to having the same label of low coherence state. Our choice of threshold therefore represents an optimal trade-off between classification accuracy and significance under the learning state permutation test.

\begin{figure}[t]
\centering
\includegraphics[width=0.5\textwidth]{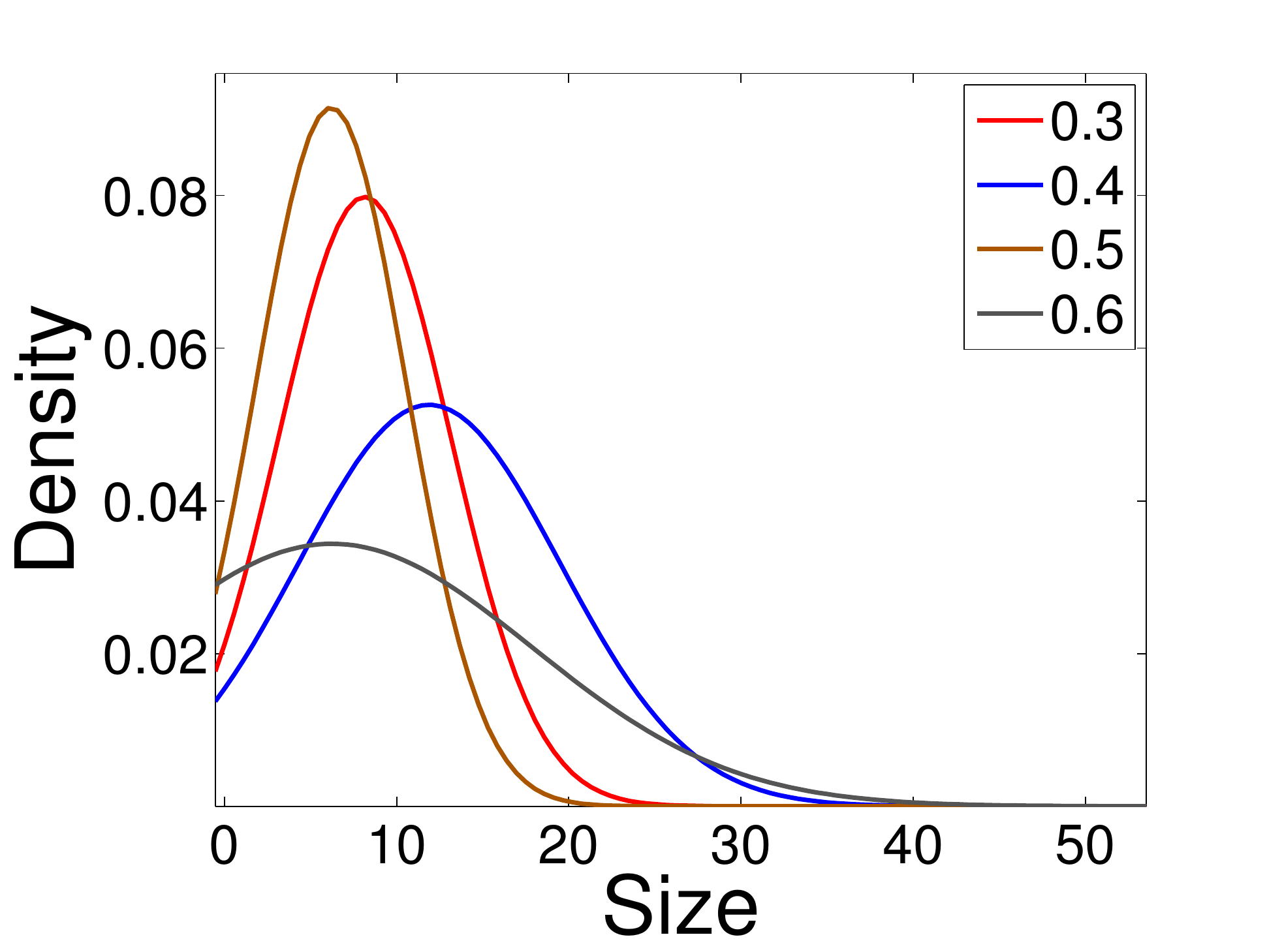}
\caption{Distribution of obtained subgraph sizes at different coherence thresholds: 0.3, 0.4, 0.5 and 0.6. 
While at thresholds $0.3$ and $0.5$, the distribution of accurate subgraphs sizes has a relatively smaller variance and subgraphs are smaller on average than for threshold values $0.4$ and $0.6$, the observed subgraphs are not significant (see Table~\ref{tab:pvalue_coherence}). A threshold of $0.4$ results in significant number of accurate subgraphs based on both permutation tests. }
\label{fig:size_threshold}
\end{figure}

\begin{table}[H]
\centering
{\scriptsize
\begin{tabular}{|c|c|c|c|c|} \hline
Coherence Threshold    & \% high coherence & \# subgraphs & LP p-value & EP p-value \\ \hline	
$0.3$           &$9$             &$73$    &$0.17$	& $1$\\\hline
$0.4$           &$3$            &$34$    &$\mathbf{0.03}$	& $\mathbf{<0.01}$\\ \hline
$0.5$           &$1$            &$15$    &$0.31$	&$1$\\ \hline
$0.6$           &$0.4$            &$4$    &$0.31$	&$1$\\ \hline
$0.7$           &$0.1$            &$0$    &N/A	&N/A\\ \hline
$0.8$           &$0$            &$0$    &N/A	&N/A\\ \hline

\end{tabular}	
}
\caption{Percentage of high coherence edges, number of discriminative subgraphs (of size at most 10) and corresponding $p$-values for varying coherence threshold. High thresholds (exceeding 0.6) result in no discriminative subgraphs. 
In the range $0.3$ to $0.6$, only at a threshold of $0.4$ the number of discriminative subgraphs are significant ($p$-values smaller than 0.03 for both learning state and edge coherence permutations) while other thresholds result in higher $p$-values.}
\label{tab:pvalue_coherence}
\end{table}

Figure~\ref{fig:size_threshold} shows the distribution of obtained accurate subgraph sizes at different thresholds. As can be seen, thresholds $0.3$ and $0.5$ result in relatively small subgraphs on average. However, the number of mined subgraphs are not significant under the LP or EP permutation tests. Table~\ref{tab:pvalue_coherence} summarizes the significance and number of small (at most 10-edge) discriminative subgraphs. At a threshold of $0.2$ (not shown in the figure and table), about half of the functional edges are deemed to be in the high coherence state, which is physiologically unlikely. Hence, we consider settings of the threshold exceeding $0.3$ that result in small ($\leq10\%$) percentage of all possible edges being in high coherence state. 

The lack of significance at lower thresholds (at approximately $0.3$) is due to the fact that each dual graph contains on average $9\%$ of its vertices in the high coherence state; permutation of vertex labels with this ratio creates very similar size distributions. At a threshold of $0.4$, we get only about $3$ percent of the edges in high coherence state and $p$-values for both permutation tests are below $0.03$. The percentage of high coherence edges drops to about $1\%$ at the threshold of approximately $0.5$, which makes the prediction of learning rate difficult; a small or nonexistent number of accurate subgraphs exist at higher thresholds because vertex coherence values are no longer discriminative.

\noindent{\bf Global Learning Rate States.} The network state mining methodology we adopt works with categorical labels of the global states, i.e. it is formalized as an instance of classification as opposed to one of regression. While estimating continuous global scores (regression) might be of interest as well, here we focus on identifying biomarkers that differentiate between sessions in which subjects progressively increase the speed of completion of the visual cues (high learning rate sessions) from those in which no acceleration is observed (low learning rate sessions). We adopt the exponential rate of decrease of the movement time as a measure of the learning rate. Movement time is the time between the first and last button press of the 12-note sequence. A larger exponential drop-off in movement time indicates that the subject is learning well, while a smaller exponential drop-off in movement time indicates that the subject is not learning as well.

The exponential drop-off parameters for the 18 subjects are listed in Table~\ref{tab:sessions}. The coloring of scores in the table are based on a threshold of $-0.06$ (which we discuss in more detail in the next paragraph) used to convert the continuous drop-off values into the categorical low versus high learning rate states. Slopes smaller than the chosen threshold correspond to high learning rate (blue) and slopes larger than the threshold correspond to low learning rate (yellow). In the first session, 13 of the subjects are ``learning'' the motor task as their rates are below the threshold, while 5 of the subjects are not learning. Because we are only interested in early learning, the 13 subjects that were labeled as being in a high learning state in the first session were discarded from the second and third session data sets. In the third session, 1 subject transitions from a low to a high learning rate state while the remaining 4 low learning rate subjects remain in the low learning rate state. Exponential drop-off parameters could not be fit to subject 16 in Session 2 and subject 1 in Session 3 due to flat or negative learning rates; these sessions were therefore excluded. Note that our estimate of learning rate is independent of how fast the the subject is at the beginning of the session, a feature which is driven by biomechanics rather than ability to learn. 

\renewcommand{\tabcolsep}{3pt}
\begin{table}[t]
\label{tab:sessions}
\centering
\scriptsize
    \begin{tabular}{|c|c|c|c|c|c|c|c|c|c|c|c|c|c|c|c|c|c|c|}
    \hline
    ~  & 1     & 2     & 3     & 4     & 5     & 6     & 7     & 8     & 9     & 10    & 11    & 12    & 13    & 14    & 15    & 16    & 17    & 18    \\ \hline
    {\bf Session 1} &\cellcolor{blue} .12 &\cellcolor{blue} .10 & \cellcolor{yellow}.05 &\cellcolor{blue} .09 &\cellcolor{blue} .09 &\cellcolor{blue} .07 & \cellcolor{yellow}.04 &\cellcolor{blue} .13 &\cellcolor{blue} .07 &\cellcolor{blue} .07 &\cellcolor{blue} .07 &\cellcolor{blue} .13 & \cellcolor{yellow}.04 &\cellcolor{blue} .08 &\cellcolor{blue} .10 & \cellcolor{yellow}0     & \cellcolor{yellow}.04 &\cellcolor{blue} .11 \\ \hline
    {\bf Session 2} & .01 & .03 & \cellcolor{yellow}.01 & .03 & .06 & .04 & \cellcolor{yellow}.03 & .05 & .02 & .04 & .05 & .03 & \cellcolor{yellow}.04 & .04 & .02 & \cellcolor{gray}NA   & \cellcolor{yellow}.04 & .03 \\ \hline
    {\bf Session 3} & \cellcolor{gray}NA& .03 & \cellcolor{yellow}0 & .1  & .07 & .05 & \cellcolor{blue}.07 & .08 & .02 & .01 & .02 & .03 & \cellcolor{yellow}.04 & .03 & .01 & \cellcolor{yellow}.05 & \cellcolor{yellow}.03 & .03 \\ \hline
    \end{tabular}
    \caption {Exponential drop-off parameters (rate of decrease) of movement time for each subject and experimental session. High-learning rate sessions are colored blue and low learning rate ones yellow. All but five subjects exhibit a high learning rate in the first session. One of those five subjects transitions into a high learning rate in their third session. Exponential drop-off of the movement time for two of the sessions (Subject 1 Session 3 and Subject 16 session 2) could not be fitted (colored gray).}
\end{table}

\begin{figure}[H]
    \centering
    \label{fig:learning_rate_stability}
    \subfigure[Stability of top regions]{
       \includegraphics[width=0.4\textwidth]{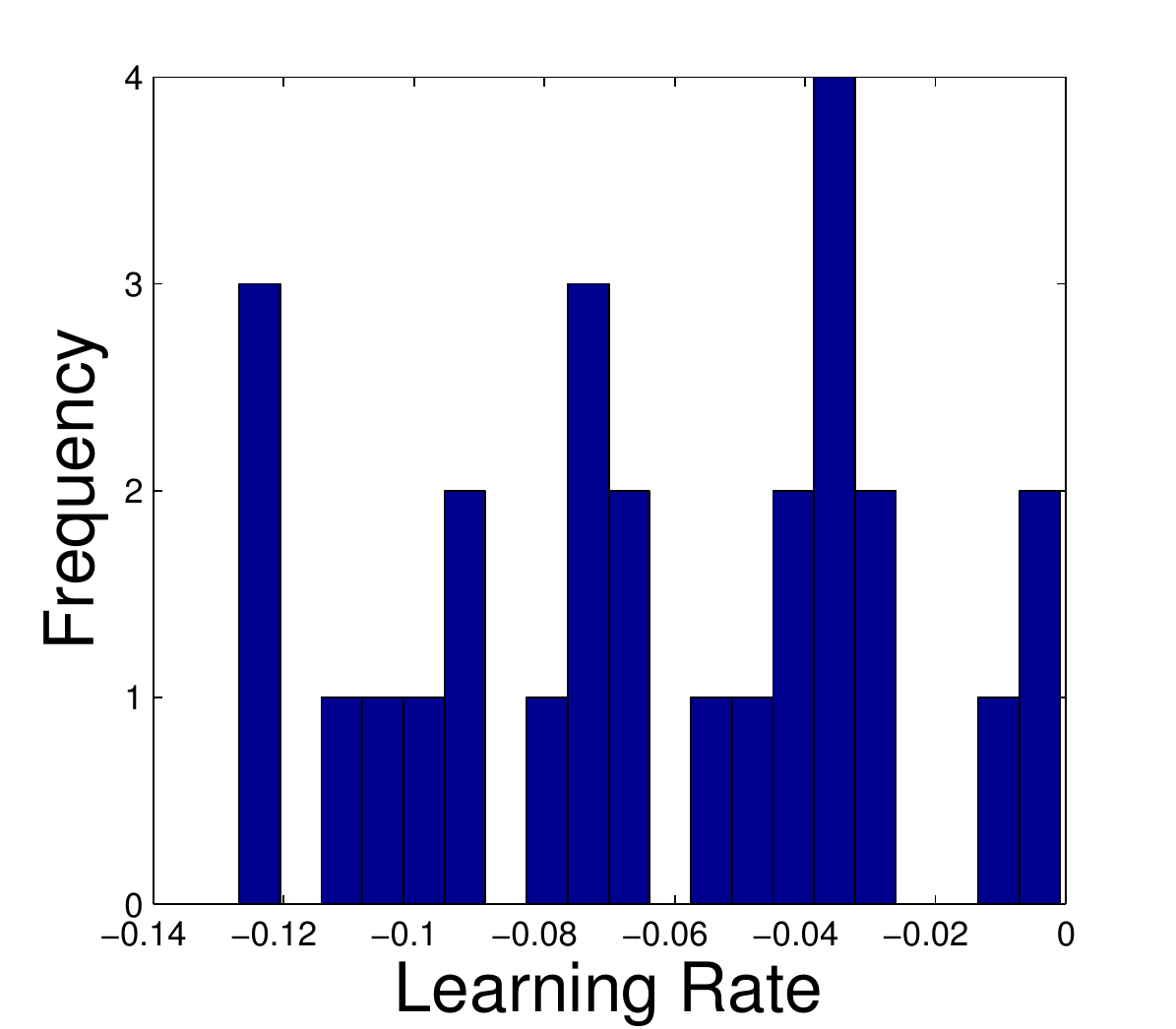}
       \label{fig:learning_rate}
    }
    \subfigure[Stability of top edges] {
       \includegraphics[width=0.4\textwidth]{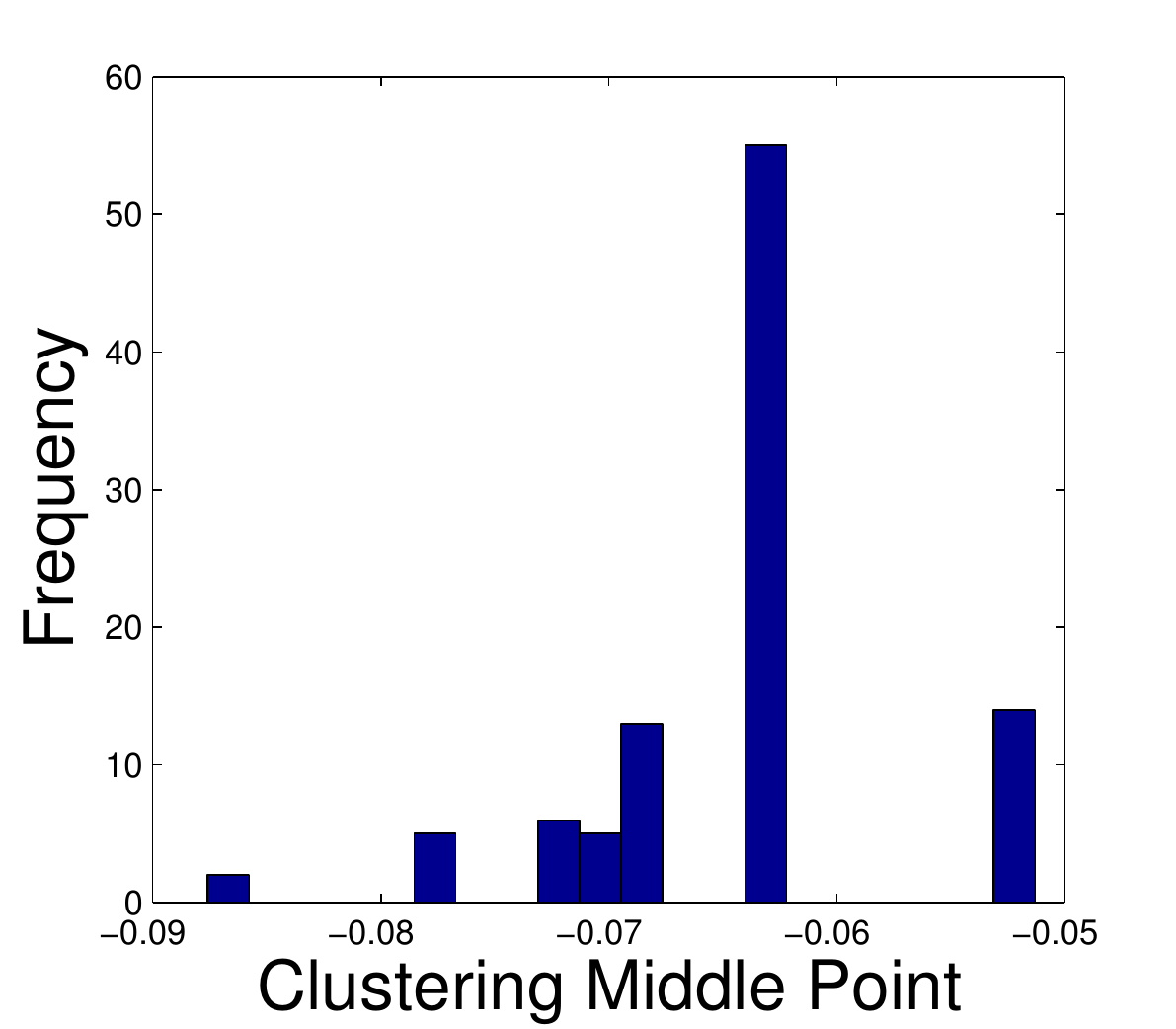}
       \label{fig:learning_rate_dist}
    }
    \caption{Distribution of the learning rate in sessions of interest~\subref{fig:learning_rate} and middle clustering point distribution~\subref{fig:learning_rate_dist} when clustering the values 100 times applying a perturbation noise of 3 standard deviations of the original values.}
\end{figure}

We selected a threshold of $-0.06$ to delineate high versus low learning rate states. To choose this threshold, we use a clustering approach with perturbations and estimate the optimal threshold between states based on the stability of multiple clustering solutions. To begin, we randomly perturbed the measured drop-off parameters from all valid sessions by adding Gaussian noise with a magnitude of 3 standard deviations of the original learning rates to each parameter value. We create 100 perturbed instances of the measured exponential drop-off parameters and clustered them using k-means clustering (k=2). We compare the different clusterings using the Jaccard index as a similarity measure. A clustering is considered ``stable'' if it is approximately maintained across multiple perturbations. The vast majority of clustering pairs have a Jaccard index of 1 (perfect similarity) and $-0.06$ is the middle point separating the points in the obtained clusterings. We show the distribution of the original learning rate values in Fig.~\ref{fig:learning_rate} and the distribution of the middle point between two perturbed clusters in Fig.~\ref{fig:learning_rate_dist}. Both figures single out $-0.06$ as an optimal threshold.

Using the selected threshold and discarding sessions preceded by high learning rate sessions of the same subject, we obtain 14 high learning rate session-specific networks and 13 low learning rate ones (color-coded with blue and yellow respectively in Tab.~\ref{tab:sessions}). We apply our discriminative biomarker approach considering all 27 networks simultaneously together with their global learning state labels.

\subsection{In-Network Discriminative Subgraph Mining}
\label{sec:mining}

Our discriminative subgraph discovery approach -- MINDS-prune -- is an extension of a recent method by Ranu and colleagues~\cite{MINDS} in which they proposed a method for network-constrained mining of decision trees for global state network label classification. In this setting, the available features for a classification problem reside on nodes in a network and the method discovers decision trees, restricted to features that form a connected subgraph in the network (network-constrained decision trees). The method's constraints could be viewed as a way to regularize the learned classifier using the inherent structure among features. The approach adopts a greedy algorithm to surpass the computational challenge of building optimal network-constrained decision trees (NCDTs). Another computational challenge is the exponential subgraph search space. To address this challenge, MINDS performs Markov Chain Monte Carlo sampling over the space of the possible subgraphs of features. The transition probabilities are based on improvement of the predictive ability of the currently selected subtree (for details refer to~\cite{MINDS}). MINDS structures the subgraph search space into an edge-weighted meta-graph where each node is a distinct subgraph, each edge is an edit (either insertion/deletion of a node to/from the subgraph) and each edge weight is the quantified impact of the edits to the accuracy. By performing a series of edits, a subgraph of the network can be transformed into any other subgraph of the network. The authors apply the method to gene-expression networks in order to find genes, whose level predicts disease states.

In our setting, we have a shared connectivity structure in the form of edge-dual graphs across networks, and vertices (features) have binary states corresponding to high and low coherence states (associated with the corresponding edges in the original graph). The network instance label is one of high versus low learning rate.  To find biomarkers that are small and accurate, we extend MINDS by a post-processing pruning phase. First, we execute MINDS and then we further compact the obtained decision trees by greedily excluding edges while maintaining connectivity and the accuracy of the originally sampled subgraph. This process is applied to every tree independently and only the smallest accurate trees are maintained. We term this extended method MINDS-prune.

\noindent{\bf Testing the Generalization Properties of MINDS-Prune.} The biomarkers we report in the previous sections were obtained by applying MINDS-prune to the full set of 27 networks. Biomarkers of training accuracy exceeding $0.9$ were kept. While these are the best biomarkers that make use of all of the training data, here we test the generalization properties of our technique using classification cross-validation. Cross validation is a common way of testing if a classifier overfits to the presented training data. As part of the cross-validation test, the training instances are partitioned in $k$ folds (subsets). A classifier is build based on $k-1$ of the folds (training) and its accuracy is then tested on the left-out fold (testing). The same procedure is repeated using each of the $k$ folds as testing. Naturally, the average accuracy in cross-validation is lower than the training accuracy when using all the annotated data. However, this test is useful to compare a classifier to baselines and to results of a random classification.

To test MINDS-prune's generalization properties, we perform a 9-fold cross validation. A common setting for the number of folds is 10, however, we chose to perform 9-fold validation as our number of instances is small and 9 is a multiple of the total instances (27 session-specific global state networks). We compute the subgraphs and their corresponding decision trees by leaving one of the folds (3 instances) out and using the other 8 folds (24 instances) as training.  We then test the accuracy of the obtained biomarker decision trees based on training on the left out testing instance.

For a relative baseline, we compare the testing accuracy of MINDS-prune with that of an SVM classifier that works with the same instances, but is oblivious to the network structure. All $6216$ features are given to the SVM while performing cross validation. Both SVM and MINDS-prune methods have the same testing accuracy of $0.69$ on average. Importantly, our method MINDS-prune performs as well as SVM while using only a small subset of the features (i.e., those included in the discriminative subgraphs), while SVM uses all features. MINDS-prune manages to select the most discriminative features among all features while taking into account the connected structure of features as well. Moreover, it also provides relevant domain insights: the subgraphs identified correspond to connected functional edges relating to regions in the motor and visual systems known to be important in motor learning.

One question of interests is whether the vertices (corresponding to functional edges) included in high-accuracy subgraphs agree across folds. To measure this, we focus our attention on subgraphs that in testing (i.e. when classifying the left-out fold) classify at least 2 out of the 3 testing instances correctly ( i.e. $66\%$ testing accuracy). Note that lower accuracy options (one or zero out of three correct predictions) would correspond to worse than random classification performance, which for two balanced classes is $50\%$. 

\begin{figure}[H]
\centering
\includegraphics[width=0.7\textwidth]{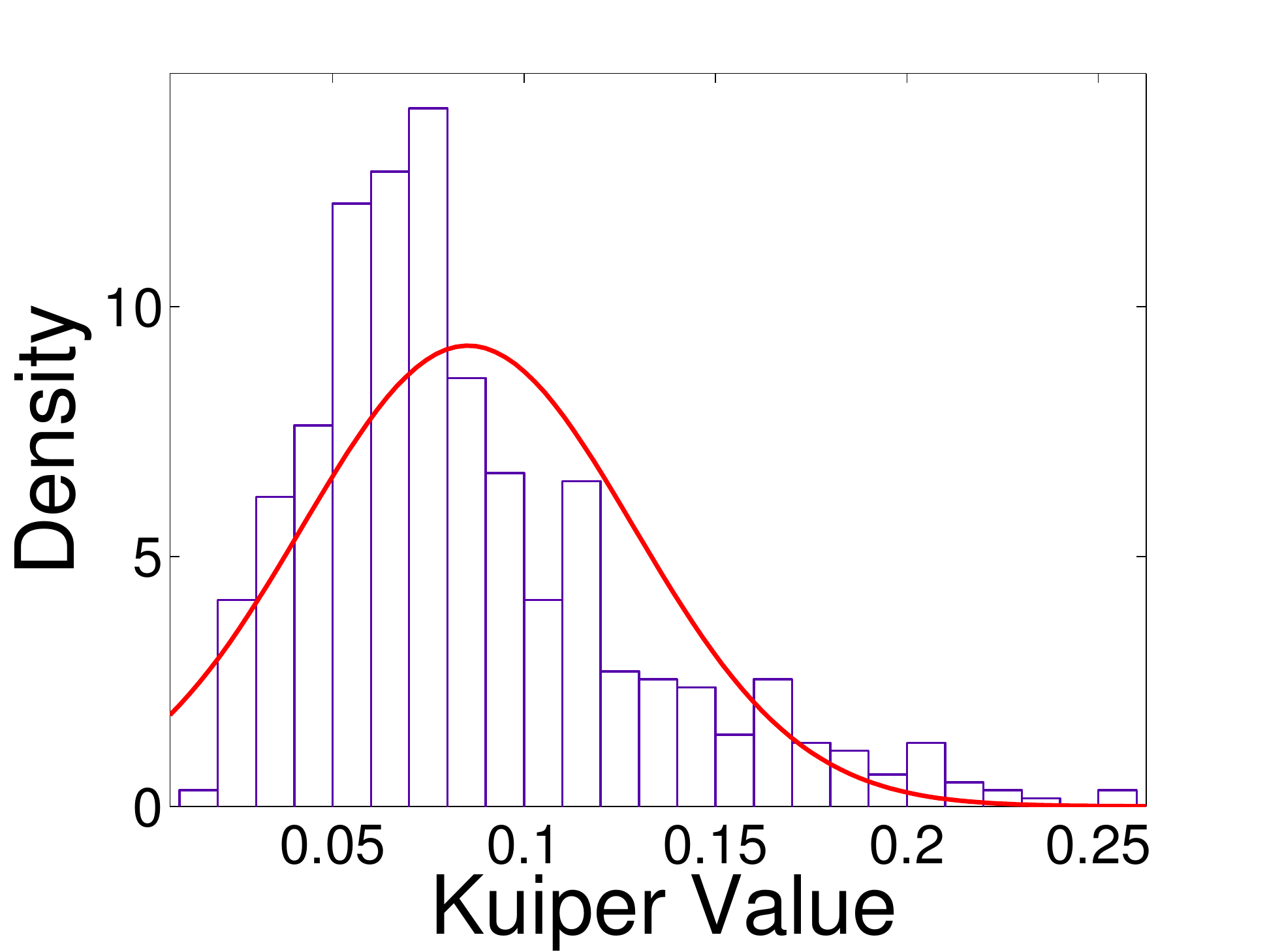}
\caption{Kuiper Statistics for pairwise subnetwork similarity distributions of 9 cross validation folds. The Kuiper statistic frequency is plotted as a function of the Kuiper values obtained from pairwise comparisons of subgraph distributions in pairs of testing folds. A Kuiper statistic of 0 corresponds to similar subgraphs within a pair of folds, while a value of 1 corresponds to distinct graphs. The average observed pairwise Kuiper statistic value is close to 0, demonstrating stability in the mined subgraphs.}
\label{fig:kuiper_folds}
\end{figure}

We compare the similarity of obtained accurate-in-testing subgraphs in the 9 folds to analyze their consistency. For this comparison, we compute pairwise subgraph similarities in each pair of folds. Subgraph similarity is defined as the fraction of overlapping nodes between any two subgraphs. We compute the pairwise subgraph similarity distribution of each fold pair (36 pairs from 9 folds), and use the Kuiper statistic to quantify the agreement of these distributions. The Kuiper statistic will be 0 if the distributions are the same and 1 if the distributions are distinct. A plot of the Kuiper statistic distribution is presented in Figure~\ref{fig:kuiper_folds}. The mean of the Kuiper statistic is close to 0 which indicates a significant similarity of subgraphs generated in each fold.

\begin{table}[h]
\centering
{\scriptsize
\begin{tabular}{|l|c|c|c|c|c|} \hline
Fold \# & Top 1 Node & Top 2 Node & Top 3 Node  \\ \hline	
$1$ & L Precentral gyrus & L Precentral gyrus & L Occipital fusiform gyrus\\
& L Occipital fusiform gyrus & L Supplemental motor area & R Occipital fusiform gyrus\\ \hline
$2$ & L Occipital fusiform gyrus & R Precentral gyrus & L Intracalcarine cortex \\
& L Occipital pole & R Lingual gyrus & L Occipital fusiform gyrus \\ \hline
$3$ & L Insular cortex & R Lingual gyrus & L Lingual gyrus\\
& R Insular cortex & R Occipital pole & R Occipital pole\\ \hline
$4$ & L Lingual gyrus & L Occipital fusiform gyrus & L Lateral occipital cortex inferior \\
& R Occipital fusiform gyrus & L Occipital pole & R Occipital fusiform gyrus \\ \hline
$5$ & R Heschl's gyrus & R Supplemental motor area & R Planum polare\\
& R Planum temporale & R Planum temporale & R Heschl's gyrus \\ \hline
$6$ & L Occipital fusiform gyrus & L Lateral occipital cortex inferior & L Cuneus cortex \\
& R Occipital fusiform gyrus & R Occipital fusiform gyrus & L Occipital fusiform gyrus\\ \hline
$7$ & L Occipital fusiform gyrus & L Supplemental motor area & L Supplemental motor area\\
& R Occipital fusiform gyrus & R Intracalcarine cortex & L Occipital fusiform gyrus \\ \hline
$8$ & L Occipital fusiform gyrus & L Occipital pole & L Precentral gyrus \\
& L Occipital pole & R Cuneus cortex & R Postcentral gyrus \\ \hline
$9$ & R Occipital fusiform gyrus & L Intracalcarine cortex & L Intracalcarine cortex \\
& R Occipital pole & L Occipital fusiform gyrus & R Occipital pole \\ \hline
\end{tabular}	
}
\caption{Top three most frequently identified functional edges in each of the nine folds for cross validation. While there is some variation across folds, due to the relatively small number of training instances and large feature space, some identified areas including the occipital fusiform gyrus and the occipital pole agree across folds.}
\label{tab:top3nodes}
\end{table}

Moreover, different folds share common cortical regions. Table~\ref{tab:top3nodes} presents the top 3 frequently identified functional edges (vertices in the edge-dual graph) from each fold. While there is some variation across folds, due to the relatively small number of training instances and large feature space, some identified areas agree across folds. For example, the occipital fusiform gyrus is frequent in 7 out of the 9 folds and the occipital pole in 5 out of the 9 folds. While the frequent regions within subgraphs when using all instances for training (reported in Section 2) are all identified as frequent in some of the folds, we do not expect a perfect agreement since in cross validation our method works with fewer training instances. The important consequence of the cross-validation results is that our method retains testing accuracy measurable with that of state-of-the-art complex classifiers such as SVM and higher than the accuracy expected at random. Our method has the advantage of incorporating connectivity among the features that in turn enables interpretation of the discovered biomarkers.

\subsection{Sensitivity to Parameters}

Our subgraph mining approach MINDS-prune has several parameters that might affect the quality of obtained results. Next, we test and discuss the sensitivity to some of those parameters and our approach to selecting them.

\begin{figure}[t]
    \centering
    \label{fig:seed_size_stability}
    \subfigure[Stability of top regions]{
       \includegraphics[width=0.4\textwidth]{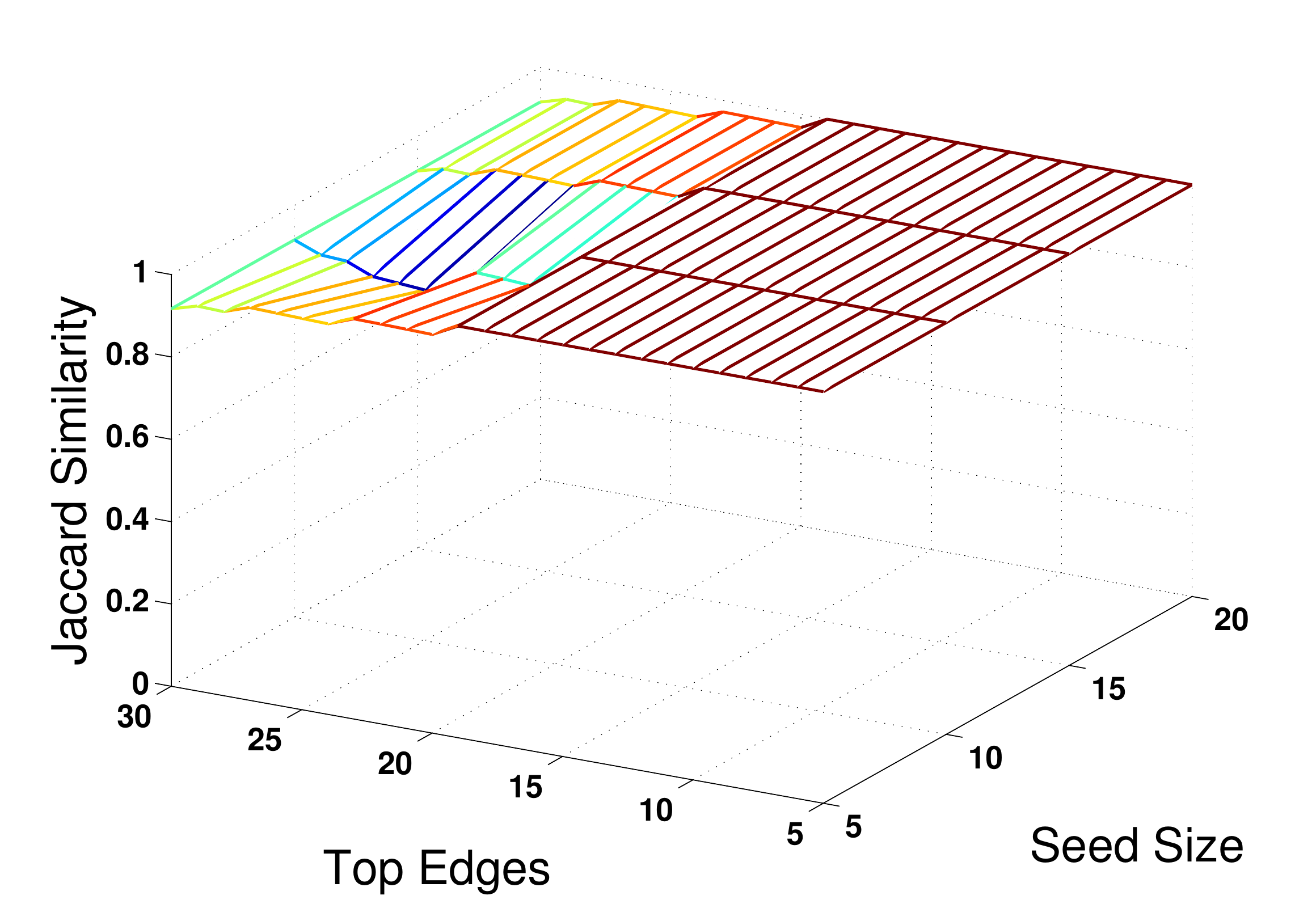}
       \label{fig:edges_seed}
    }
    \subfigure[Stability of top edges] {
       \includegraphics[width=0.4\textwidth]{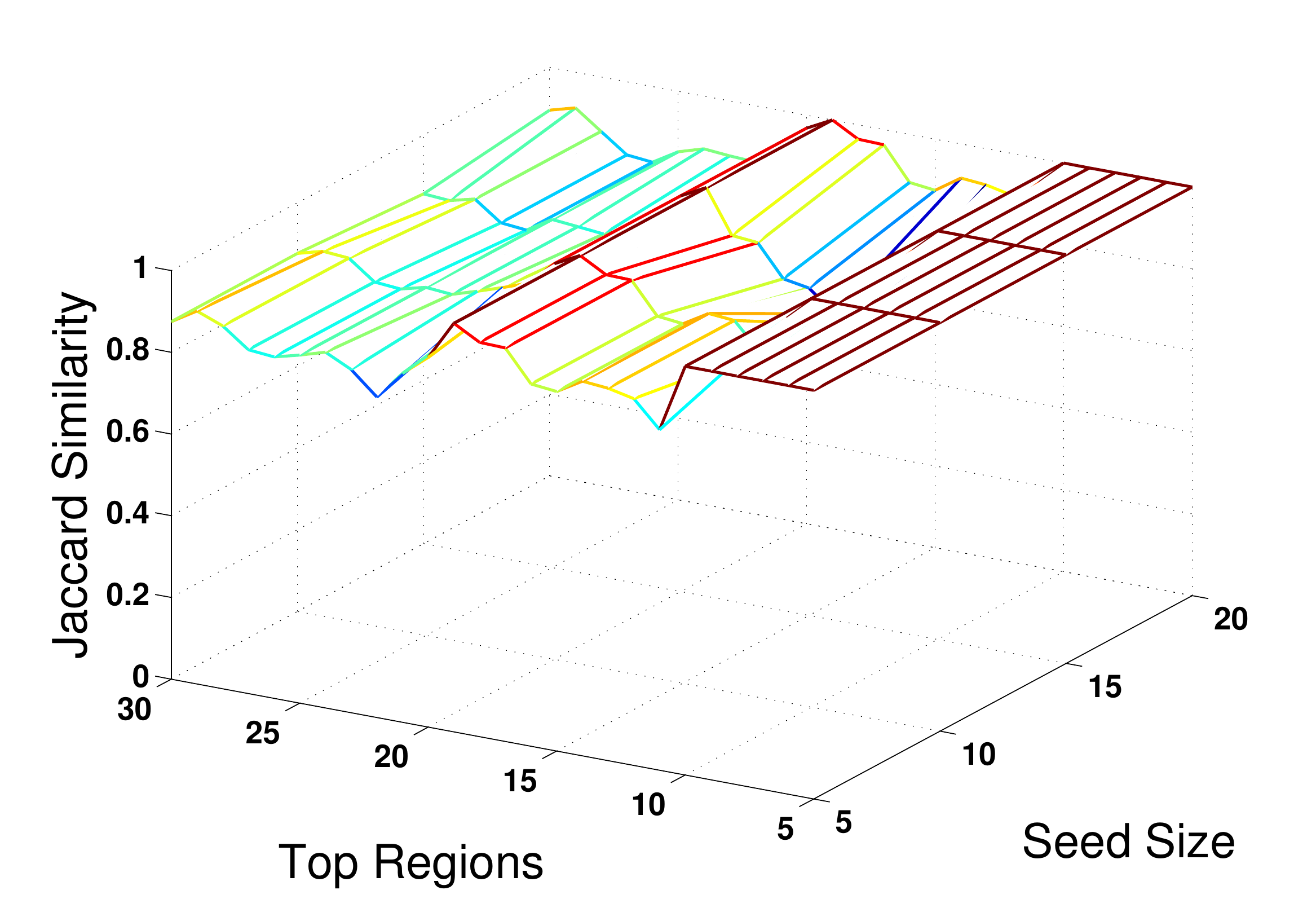}
       \label{fig:regions_seed}
    }
    \caption{Jaccard similarities of top k edge~\subref{fig:edges_seed} and region~\subref{fig:regions_seed} lists for different subgraph seed sizes of 20-edge, 15-edge, 10-edge, and 5-edge.}
\end{figure}

As discussed earlier we perform sampling runs starting from seed subgraphs of size 10-edges and retain only those that result in good classification accuracy.  Here we vary this initial seed size, starting from a 20-edge seed subgraph and perform separate independent sampling runs with random subgraphs of size 15-edge, 10-edge, and 5-edge. We measure the stability of the most frequent regions and edges obtained from the runs based on different seed sizes. The top region and edge lists for decreasing subgraph seeds are compared to all other seed sizes using Jaccard similarity and the average similarity (to different sizes and over multiple instantiations of the original 20-edge seed) is presented in Figure~\ref{fig:seed_size_stability}. Both top regions and edges are stable when starting with seed subgraphs of different sizes. The list of most frequently occurring edges/regions have average similarity of close to 1 (unchanged), while extending the list to the top 10, 20 and 30 slightly decreases this average similarity, retaining similarity values of higher than $0.6$.

In our MCMC sampling, we use a ``burn-in'' period of $1000$ moves during which samples are discarded (not added to the sample set). While this is a common practice to reduce the dependence on the initial state, there exists no rigorous theoretical proof of advantages/disadvantages of burn-in period in MCMC sampling. Instead this issue is typically left as a choice for the practitioner. Our experiments have shown that top regions and edges tend to be more stable if burn-in is used. At the same time larger burn-in has computational costs and as a result we set our burn-in to half of the actual sampling iterations. We draw $2000$ actual sample subgraphs and apply a post-processing phase that retains only high-accuracy and minimal subgraphs. While a larger number of sample subgraphs results in denser sampling, the set of those that survive the post-processing accuracy filtering stabilizes at about $2000$ samples. Similarly, the set of the top most frequent regions and the set of the top most frequent edges also stabilizes at about $2000$ samples.

\section{Conclusions}

We developed a general approach for discovery of brain activity biomarkers from fMRI data and applied it to data from sensorimotor task learning. Using data from an fMRI study coupled with performance (rate of learning within a session), we demonstrate the existence of subgraphs whose edge states are discriminative and thus can predict the rate of learning within the task. Analysis of the cortical regions and functional edges involved in the most predictive subnetworks demonstrate that the regions we obtain are both statistically significant and domain-relevant. While we focus on data from a learning experiment, our framework can also be applied to other cognitive tasks, and can further be used to identify biomarkers specific to neurological and psychiatric disease.


\bibliographystyle{abbrv}


\section*{Appendix}


\section*{Supplementary material}

List of Brain Regions

| | | |
|---|---|---|
| 1 | L | Frontal pole |
| 2 | L | Insular cortex |
| 3 | L | Superior frontal gyrus |
| 4 | L | Middle frontal gyrus |
| 5 | L | Inferior frontal gyrus, pars triangularis |
| 6 | L | Inferior frontal gyrus, pars opercularis |
| 7 | L | Precentral gyrus |
| 8 | L | Temporal pole |
| 9 | L | Superior temporal gyrus, anterior |
| 10 | L | Superior temporal gyrus, posterior |
| 11 | L | Middle temporal gyrus, anterior |
| 12 | L | Middle temporal gyrus, posterior |
| 13 | L | Middle temporal gyrus, temporooccipital |
| 14 | L | Inferior temporal gyrus, anterior |
| 15 | L | Inferior temporal gyrus, posterior |
| 16 | L | Inferior temporal gyrus, temporooccipital |
| 17 | L | Postcentral gyrus |
| 18 | L | Superior parietal lobule |
| 19 | L | Supramarginal gyrus, anterior |
| 20 | L | Supramarginal gyrus, posterior |
| 21 | L | Angular gyrus |
| 22 | L | Lateral occipital cortex, superior |
| 23 | L | Lateral occipital cortex, inferior |
| 24 | L | Intracalcarine cortex |
| 25 | L | Frontal medial cortex |
| 26 | L | Supplemental motor area |
| 27 | L | Subcallosal cortex |
| 28 | L | Paracingulate gyrus |
| 29 | L | Cingulate gyrus, anterior |
| 30 | L | Cingulate gyrus, posterior |
| 31 | L | Precuneus cortex |
| 32 | L | Cuneus cortex |
| 33 | L | Orbital frontal cortex |
| 34 | L | Parahippocampal gyrus, anterior |
| 35 | L | Parahippocampal gyrus, posterior |
| 36 | L | Lingual gyrus |
| 37 | L | Temporal fusiform cortex, anterior |
| 38 | L | Temporal fusiform cortex, posterior |
| 39 | L | Temporal occipital fusiform cortex |
| 40 | L | Occipital fusiform gyrus |
| 41 | L | Frontal operculum cortex |
| 42 | L | Central opercular cortex |
| 43 | L | Parietal operculum cortex |
| 44 | L | Planum polare |
| 45 | L | Heschl's gyrus |
| 46 | L | Planum temporale |
| 47 | L | Supercalcarine cortex |

| | | |
|---|---|---|
| 48 | L | Occipital pole |
| 49 | R | Frontal pole |
| 50 | R | Insular cortex |
| 51 | R | Superior frontal gyrus |
| 52 | R | Middle frontal gyrus |
| 53 | R | Inferior frontal gyrus, pars triangularis |
| 54 | R | Inferior frontal gyrus, pars opercularis |
| 55 | R | Precentral gyrus |
| 56 | R | Temporal pole |
| 57 | R | Superior temporal gyrus, anterior |
| 58 | R | Superior temporal gyrus, posterior |
| 59 | R | Middle temporal gyrus, anterior |
| 60 | R | Middle temporal gyrus, posterior |
| 61 | R | Middle temporal gyrus, temporooccipital |
| 62 | R | Inferior temporal gyrus, anterior |
| 63 | R | Inferior temporal gyrus, posterior |
| 64 | R | Inferior temporal gyrus, temporooccipital |
| 65 | R | Postcentral gyrus |
| 66 | R | Superior parietal lobule |
| 67 | R | Supramarginal gyrus, anterior |
| 68 | R | Supramarginal gyrus, posterior |
| 69 | R | Angular gyrus |
| 70 | R | Lateral occipital cortex, superior |
| 71 | R | Lateral occipital cortex, inferior |
| 72 | R | Intracalcarine cortex |
| 73 | R | Frontal medial cortex |
| 74 | R | Supplemental motor area |
| 75 | R | Subcallosal cortex |
| 76 | R | Paracingulate gyrus |
| 77 | R | Cingulate gyrus, anterior |
| 78 | R | Cingulate gyrus, posterior |
| 79 | R | Precuneus cortex |
| 80 | R | Cuneus cortex |
| 81 | R | Orbital frontal cortex |
| 82 | R | Parahippocampal gyrus, anterior |
| 83 | R | Parahippocampal gyrus, posterior |
| 84 | R | Lingual gyrus |
| 85 | R | Temporal fusiform cortex, anterior |
| 86 | R | Temporal fusiform cortex, posterior |
| 87 | R | Temporal occipital fusiform cortex |
| 88 | R | Occipital fusiform gyrus |
| 89 | R | Fronal operculum cortex |
| 90 | R | Central opercular cortex |
| 91 | R | Parietal operculum cortex |
| 92 | R | Planum polare |
| 93 | R | Heschl's gyrus |
| 94 | R | Planum temporale |
| 95 | R | Supercalcarine cortex |
| 96 | R | Occipital pole |

| | | |
|---|---|---|
| 97 | L | Caudate |
| 98 | L | Putamen |
| 99 | L | Globus pallidus |
| 100 | L | Thalamus |
| 101 | L | Nucleus Accumbens |
| 102 | L | Parahippocampal gyrus (superior to ROIs 34,35) |
| 103 | L | Hippocampus |
| 104 | L | Brainstem |
| 105 | R | Caudate |
| 106 | R | Putamen |
| 107 | R | Globus pallidus |
| 108 | R | Thalamus |
| 109 | R | Nucleus Accumbens |
| 110 | R | Parahippocampal gyrus (superior to ROIs 34,35) |
| 111 | R | Hippocampus |
| 112 | R | Brainstem |



\end{document}